# Pulsed heterodyne detection enables fiber-compatible, high-specificity Brillouin biomechanics in intact tissue and the living brain

Zixuan Du[1,7,†], Shuai Yao[1,†], Yun Qi[1,*], Jinrui Zhang[1], Junlin You[1], Xiaoyang Liu[1], Yangxuan Liu[2], Yun Luo[3], Ting Mi[4], Yuan Wang[5], Guanyi Zhu[1], Fei He[1], Jingjing Xie[4], Shiqing Cai[3], Weibing Yang[2], Zhisheng Yang[6,*], Long Zhang[1,*], Weibiao Chen[1,*], Fan Yang[1,7,*]

[1] Center for Biomedical Photonics, Shanghai Institute of Optics and Fine Mechanics, Chinese Academy of Sciences, Shanghai, China.
[2] Institute of Plant Physiology and Ecology, CAS Centre for Excellence in Molecular Plant Sciences, Chinese Academy of Sciences, Shanghai, China.
[3] Institute of Neuroscience, CAS Center for Excellence in Brain Science and Intelligence Technology, Chinese Academy of Sciences, Shanghai, China.
[4] State Key Laboratory of Advanced Medical Materials and Devices, ShanghaiTech University, Shanghai, China.
[5] Shanghai Sixth People's Hospital, Shanghai Jiao Tong University School of Medicine, Shanghai, China.
[6] State Key Laboratory of Information Photonics & Optical Communications, Beijing University of Posts and Telecommunications, Beijing, China.
[7] University of Chinese Academy of Sciences, Beijing, China.

[†] These authors contributed equally to this work.
Correspondence should be addressed to Y.Q. (qiyun@siom.ac.cn), Z.Y. (zhisheng.yang@bupt.edu.cn), L.Z. (lzhang@siom.ac.cn), W.C. (wbchen@siom.ac.cn) or F.Y. (yang@siom.ac.cn).

**Brillouin microscopy provides label-free, three-dimensional mechanical characterization of biological specimens, but current dispersive detection imposes two limits: the fiber background folds onto the sample spectrum, precluding single-fiber operation, and a ~250 MHz dispersion-induced instrumental broadening blurs mechanically distinct components within a focal volume. Here, we introduce pulsed heterodyne Brillouin detection (PHBD), which retrieves the spectrum electronically from temporal beat notes, overcoming both limitations. The fiber background beats outside the detection band and is rejected; elimination of dispersive broadening yields 25-MHz spectrometer resolution. Pulsed excitation reaches Brillouin-signal-shot-noise-limited detection, attaining 9.0-MHz shift precision in 3 ms at 30 mW in water with 53-fold improvement in energy efficiency over continuous-wave excitation. Through a bare 125-µm fiber, PHBD resolves regional contrast along a 4-mm insertion track in the living mouse brain; in free space, it resolves distinct Brillouin components from the cell wall and adjacent cytoplasm in strongly scattering *Arabidopsis* root tips with epi-mode.**

The mechanical properties of cells and tissues govern processes from development and regeneration to disease progression, making quantification of viscoelastic parameters such as stiffness and viscosity central to mechanobiology [1–4]. Most established assays, however, apply a mechanical load through physical contact and therefore probe accessible surfaces, or lack three-dimensional cellular resolution [5,6].

Brillouin microscopy measures mechanical properties in a label-free, three-dimensional manner at subcellular resolution without externally applied deformation. The Brillouin frequency shift and linewidth probe the elastic and viscous components, respectively, of the longitudinal mechanical response at gigahertz frequencies, and the technique has been applied across biological systems from single cells [7–10] and embryos [11–13] to spheroids and organoids [14–16], small organisms [17–19], plant tissues [20,21] and the cornea [22–24]. Current implementations rely on spontaneous Brillouin scattering typically resolved by a virtually imaged phased array (VIPA) [25–29], which employs optical dispersion and thus limits biological utility in two ways.

The first limitation concerns physical reach. Free-space configurations restrict interrogation depth typically to a few hundred micrometers, set by the objective working distance and ballistic-photon budget. A single-fiber probe delivering excitation and collecting scattered light could overcome this constraint and enable access through an endoscope, biopsy needle or catheter. However, owing to the finite free spectral range of dispersive spectrometers, Brillouin scattering generated along the delivery fiber which is more than three orders of magnitude stronger than the biological signal folds onto the sample spectrum and saturates the detector. Hollow-core fibers [30], dual-fiber probes [31] and gold-coated fibers [32,33] have demonstrated fiber-based Brillouin acquisition, though integration times of 2.5–30 s per point have constrained application in living specimens.

The second limitation concerns mechanical specificity, defined as the ability to attribute the measured Brillouin spectrum to a specific mechanical constituent rather than to an unresolved mixture. This specificity depends on the spectral resolution of the spectrometer. With a typical instrumental linewidth of ~250 MHz [14,25,34], dispersive spectrometers merge mechanically distinct constituents sharing a focal volume into a single broadened peak. A single-peak fit then returns a shift and linewidth that reflect the volume-weighted average of the mixture, and subvoxel heterogeneity manifests as apparent linewidth broadening rather than being recognized as coexisting components. Two approaches have surpassed this limit, but each sacrifices a capability important for biological application. Stimulated Brillouin scattering (SBS) microscopy replaces dispersion with a pump-probe scheme [16,19,35–39], yet requires a dual-objective transmission geometry, which precludes single-ended detection and restricts measurements to semi-transparent samples ~100 µm thick [16,19,40]. Continuous-wave (CW) heterodyne detection [41] attains high spectral resolution with single-ended access, but requires high optical power (~276 mW) and long pixel dwell times (~100 ms), yielding substantially higher optical doses. High spectral resolution at a biologically compatible dose in a single-ended geometry has therefore remained challenging.

Here we introduce pulsed heterodyne Brillouin detection (PHBD), which overcomes both limitations by resolving the Brillouin spectrum electronically, from temporal beat notes, rather than by optical dispersion. Because fused silica has a Brillouin shift of ~22 GHz at 780 nm, ~17 GHz above that of soft tissue, the fiber background beats far outside the electronic detection band and is discarded; retrieving the spectrum in the time domain simultaneously reduces instrumental broadening, giving 25-MHz spectrometer resolution, tenfold finer than a VIPA spectrometer. Pulsed excitation reaches Brillouin-signal-shot-noise-limited detection, attaining 9.0-MHz shift precision in 3 ms at 30 mW with 53-fold improvement in energy-efficiency over continuous-wave excitation. Fiber-coupled PHBD acquires spectra two to three orders of magnitude faster than existing fiber probes, enabling depth-resolved Brillouin spectroscopy along a ~4-mm insertion track in the living mouse brain and regional comparison in the porcine stomach approached through the lumen. In free space, it resolves components as close as 240 MHz apart in epi-mode and enables Brillouin imaging at $0.6 \times 0.6 \times 2.4$ µm$^3$ resolution in strongly scattering tissues, including living *Arabidopsis* root tips and freshly excised porcine renal cortex.

## Results

### Principle of pulsed heterodyne Brillouin detection

PHBD retrieves the Brillouin spectrum in the time domain rather than by optical dispersion (Fig. 1a). Excitation light generates spontaneous Brillouin-scattered signal in the sample, and the backscattered light is recombined with a local oscillator (LO). Because signal and LO differ in optical frequency, their superposition produces a beat note at the difference frequency, decaying at the rate of acoustic damping. Fourier analysis of this transient yields the Brillouin spectrum, from which we extract the shift $\nu_B$, the linewidth $\gamma_B$ and the intensity $I_B$ (Fig. 1a, top).

The LO detuning sets the frequency layout of the measurement (Fig. 1a, bottom). With the LO at $\nu_{LO}$, Stokes Brillouin light at $\nu_P - \nu_B$ is down-converted to $\Delta\nu = \nu_P - \nu_B - \nu_{LO}$, whereas elastically scattered excitation at $\nu_P$ appears at $\nu_P - \nu_{LO}$. Choosing $\Delta\nu$ places the sample Brillouin beat note within the detection bandwidth whereas the fiber background from fused silica, whose Brillouin shift is approximately 17 GHz higher than that of soft tissue, produces a beat note beyond the electronic cutoff and is therefore never digitized. Removing the dispersive element simultaneously removes its instrumental broadening, so Brillouin components separated by a few hundred megahertz remain resolvable (Supplementary Fig. 1).

The pulse enhancement follows from how signal and noise scale with power. In the constant-noise-limited regime, the noise floor is set predominantly by the LO and is therefore independent of the excitation power (Supplementary Figs. 2 and 3), whereas the Brillouin heterodyne signal grows linearly with it. Consequently, the signal-to-noise ratio (SNR) scales as $P_{exc}\sqrt{T_{exc}}$, where $P_{exc}$ is the excitation power and $T_{exc}$ is the effective excitation duration (Eq. (27) in Supplementary Note 1). Light is thus spent more efficiently at high power than over long exposures: raising $P_{exc}$ 53-fold shortens the exposure needed for a given SNR by $53^2$, or ~2,800-fold, and therefore lowers the total optical dose 53-fold (Fig. 1b, top).

Pulsed excitation realizes this high-power, short-duration operating point while keeping the average power low. The required energy is delivered as a pulse train across the acquisition window, retaining the peak power while reducing the average (Fig. 1b, bottom). Thus, the average power required to maintain a fixed SNR is reduced by the peak-power enhancement factor $Q$ since SNR scales as $\sqrt{P_{ave}Q}$ (Eq. (29) in Supplementary Note 1). The maximum pulse-enhancement benefit is reached once Brillouin-signal shot noise dominates, beyond which SNR depends only on the total delivered optical energy.

We implemented PHBD with a pulsed excitation laser and a CW LO (Supplementary Fig. 4 and Methods). Two configurations share the same heterodyne spectrometer (Fig. 1c, full optical schematics in Supplementary Figs. 5a and 6). In the fiber-compatible configuration (Fig. 1c, left), we route the excitation through a fiber circulator and deliver it by a flexible bare fiber that also collects the backscattered signal; in the microscopy configuration (Fig. 1c, middle), we focus it through a 0.7-NA objective, giving a spatial resolution of $0.6 \times 0.6 \times 2.4\ \mu m^3$ (Supplementary Fig. 7, Supplementary Note 7 and Methods). In both, we combine the collected light with the LO in a fiber coupler, detect it by a balanced photodetector and digitize the signal for spectral reconstruction (Fig. 1c, right; Methods, Supplementary Fig. 8 and Supplementary Note 4).

We validated the pulse-enhancement mechanism in water at a fixed 10-ms acquisition time (Fig. 1d). Raising the peak power from 3.5 W to 14 W while lowering the repetition rate from 640 kHz to 40 kHz, thereby reducing the average power from 80 mW to 20 mW, maintained the shift precision at ~8 MHz. Beyond this range the precision degraded as predicted, marking the onset of the Brillouin-signal-shot-noise-limited regime (Supplementary Fig. 9a and Supplementary Note 2). At the lowest average power tested (28 W peak, 10 mW average), PHBD required 53-fold lower average power at a shift precision of 9.9 MHz, compared with ~6 MHz under CW excitation at the same acquisition time on the same instrument (Supplementary Fig. 9b). It also matched the precision of the previous CW heterodyne implementation [41] at 35-fold lower average power (Supplementary Fig. 9b), although the two systems used different wavelengths (Supplementary Note 3).

**Performance of PHBD**

We first characterized the free-space configuration on distilled water. At a 3-ms pixel dwell time and 30-mW average power on the sample (36-ns pulses at 30 kHz), PHBD reached a Brillouin-shift precision (hereafter shift precision) of 9.0 MHz and a linewidth precision of 31.6 MHz (Fig. 2a,b). The spectrometer resolution is 25 MHz, set mainly by the transform limit of the recorded transient; we confirmed this independently from the laser spectrum and by cross-validation against SBS microscopy (Supplementary Figs. 10 and 11) [39]. The measured water linewidth of 289 MHz (NA = 0.12) is dominated by intrinsic acoustic damping (Supplementary Fig. 10d), confirming that instrumental broadening is negligible.

Under fiber coupling (Fig. 2c), backward-propagating Brillouin scattering from the delivery fiber saturated the VIPA camera and masked the spectral region containing the sample signal (Fig. 2d and Supplementary Fig. 12). PHBD rejected this background entirely: we acquired clean spectra at 8.9-MHz shift precision with 20-mW average power and a 10-ms dwell time (Fig. 2e and Supplementary Fig. 5a–e), matching the free-space precision through the 0.5-m delivery fiber used in this configuration.

To test whether this resolution translates into component specificity, we benchmarked PHBD against a VIPA spectrometer built following ref. 42, using a phantom in which 1% agarose and 6% NaCl solution meet at a planar interface (Fig. 2f). When we placed the focal volume on the interface, the VIPA system returned a single broadened peak (Fig. 2g), whereas PHBD resolved two constituents 240 MHz apart, at 5.09 and 5.33 GHz (Fig. 2h). PHBD therefore preserves component-specific information in heterogeneous focal volumes, whereas dispersive broadening yields a merged response that may not represent either constituent.

Frequency stability determines whether measurements can be compared across a session (Supplementary Fig. 13). Over 110 min of free-running operation, the VIPA system drifted by ~40 MHz, chiefly from the temperature sensitivity of VIPA dispersion. PHBD drifted by ~2 MHz over the same interval, without any calibration channel in the instrument.

**Fiber-coupled Brillouin spectroscopy in the living brain and the intact stomach**

We next tested whether the fiber-coupled channel could reach interrogation points millimeters beyond the range of a free-space geometry, using two modes of access. For direct insertion, we performed depth-resolved Brillouin spectroscopy along an insertion track in the mouse brain *in vivo* ($n$ = 3 mice; Fig. 3a). We advanced a single 125-µm bare optical fiber, serving simultaneously as excitation and collection channel, from the cortical surface at AP −2.06 mm, ML −1.5 mm in 20-µm steps to a depth of ~4 mm (Fig. 3a,b). The Brillouin shift rose sharply at the saline–cortex boundary, registering the fiber tip against a known landmark, and thereafter tracked the anatomy expected at these coordinates (Fig. 3c,d). The thalamus returned a higher shift than the hippocampus (mean 5.51 GHz versus 5.45 GHz), and we cleanly resolved spectra from saline, cortex, hippocampus and thalamus through the fiber (Fig. 3e). These deep structures (e.g. the thalamus, a relay hub implicated in neurodegenerative and psychiatric disorders) lie well beyond the reach of free-space Brillouin optics, and their mechanical characterization has previously relied on excised or fixed tissue. The fiber-coupled geometry removes the depth constraint imposed by objective working distance: interrogation depth is set by probe advance rather than by ballistic photon transport, and the measurement is performed under intact vascular perfusion and physiological temperature.

For luminal access, we guided a bare fiber through the lumen of an intact porcine stomach *ex vivo* ($n$ = 3, Fig. 3f). The antrum exhibited a higher and concentrated shift distribution than the fundus (5.42–5.68 GHz, mean 5.54 GHz, versus 5.21–5.59 GHz, mean 5.40 GHz; Fig. 3g). We laterally mapped a dissected stomach with a custom lensed probe and reproduced the same trend in an independent geometry ($n$ = 6; Supplementary Fig. 14). The distinct Brillouin-shift distributions observed in the antrum and fundus suggest that the antral epithelium has higher longitudinal modulus than the fundic epithelium, provided that the ratio between refractive index and density does not substantially vary across these tissues, as previously shown in zebrafish [17]. The contrast is consistent with the distinct roles of the two regions: the antrum drives gastric peristalsis, whereas the fundus serves as a compliant reservoir.

**High-resolution, low-phototoxicity epi-Brillouin imaging of unsectioned scattering tissue**
The free-space configuration combines high spectral resolution with single-sided optical access, so scattering, unsectioned specimens can be imaged without transmitting light through the whole sample. We first imaged living *Arabidopsis* root tips ($n$ = 5; Fig. 4a–e), which are strongly scattering and therefore poorly suited to dual-objective transmission despite being only ~100 μm thick. Shift maps resolved the contrast between root cap and meristematic zone (mean 5.42 versus 5.29 GHz) and delineated cell walls and nuclei within individual cells (Fig. 4c,d). At the cell wall–cytoplasm boundary, PHBD returned two peaks (Fig. 4e): we interpret these as cytoplasm (5.18 GHz, L1) and cell wall (6.05 GHz, L2), consistent with the substantially higher Brillouin shift expected from a cellulose-microfibril-reinforced matrix. Because the wall is thinner than the focal volume, its signal is normally mixed with cytoplasmic contributions in the same voxel. Resolving the two recovers cell-wall-specific information and prevents subvoxel heterogeneity from being absorbed into a single-peak fit, where it would appear as an intrinsic change in linewidth. PHBD separated still closer contributions in fibroblasts, resolving components 250 MHz apart at the medium-cytoplasm boundary and 291 MHz apart at the nucleoplasm-nucleolus boundary (Supplementary Fig. 18h,i).

We then extended PHBD to freshly excised porcine renal cortex ($n$ = 2), an unsectioned specimen more than 5 mm thick (Fig. 4f–h). Transverse maps across a 300 × 280 μm² field resolved shift heterogeneity spanning ~460 MHz, revealing compact high-shift domains embedded within a lower-shift tissue background. We recovered spectra more than 70 μm beneath the tissue surface through axial sections of the renal capsule, demonstrating near-surface three-dimensional Brillouin mapping in tissue that is neither sectioned nor optically transparent.

Because pulsed excitation reaches peak powers of tens of watts, we assessed phototoxicity under the typical imaging conditions. NIH/3T3 cells that we imaged at 30 mW average power, 33 W peak power and a 10-ms dwell time showed no detectable propidium iodide fluorescence 50 min after imaging ($n$ = 20), whereas dead-cell controls fluoresced strongly (Supplementary Fig. 15a–e); *Caenorhabditis elegans* embryos that we imaged under the same conditions continued their developmental progression ($n$ = 13; Supplementary Fig. 15f–i). Embryos that we imaged under CW excitation at 55 mW with a 50-ms dwell time arrested in development ($n$ = 9; Supplementary Fig. 15j–m), confirming the photosensitivity of this model. These assays indicate no detectable acute loss of membrane integrity and no overt developmental impairment under the tested PHBD conditions.

**Discussion**

We benchmark PHBD against representative Brillouin detection schemes in Supplementary Table 1 and characterize its performance across operating conditions in Supplementary Fig. 16. Among spontaneous dispersive methods, PHBD matches the shift precision of confocal VIPA spectrometers at approximately one-third the optical energy [7,21,23] and is comparable in acquisition time, precision and excitation power to line-scanning [11,14,43] and full-field [44] configurations (Supplementary Fig. 16a–d), while providing tenfold finer spectrometer resolution and inherent fiber compatibility (Supplementary Note 6). Because its frequency axis is referenced to a difference frequency rather than to a dispersive element, PHBD achieves substantially greater frequency stability without auxiliary reference paths or electro-optic calibration, exhibiting a drift of ~2 MHz over 110 min of free-running operation, compared with ~40 MHz for a VIPA spectrometer [10,27].

The principal advantage of PHBD over stimulated Brillouin scattering (SBS) microscopy lies in its single-ended, epi-detection geometry. SBS derives its sensitivity from nonlinear pump–probe gain, which requires two counter-propagating beams in a transmission configuration [16,19,35–40], restricting measurements to semi-transparent specimens typically ~100 µm thick [16,19,40] and precluding single-ended access to scattering tissue and fiber-coupled operation. PHBD remains spontaneous, linear and single-ended: its pulse enhancement arises from higher instantaneous power at fixed average power against a LO noise floor that does not scale with excitation power, requiring no counter-propagating beam and therefore operating identically through a free-space objective or a bare fiber. In practice, PHBD acquires spectra approximately tenfold faster than pulse-chopped SBS at comparable precision and power [16,38] (Supplementary Fig. 16e); a recent pulsed-laser SBS implementation achieves notably higher speed [39] but remains confined to semi-transparent samples and provides lower spectral resolution (Supplementary Note 5). This epi-compatible, fiber-compatible architecture enabled depth-resolved spectroscopy in the living brain, luminal access to the intact stomach, and component-resolved imaging in scattering root tips and unsectioned renal cortex, none of which can be achieved with a transmission-geometry instrument.

Continuous-wave (CW) heterodyne detection [41] also retrieves the spectrum electronically in a single-ended geometry, but requires ~276 mW and ~100 ms dwell times, confining it to free-space operation. Pulsed excitation overcomes this constraint: by concentrating optical energy into short pulses against a LO noise floor independent of excitation power, PHBD improves energy efficiency 35-fold in water and 92-fold in biological samples (Supplementary Note 3), reducing average power to levels compatible with biological specimens. Existing fiber-based Brillouin probes address the fiber background through fiber engineering, including hollow-core fibers [30], dual-fiber geometries [31] and gold-coated fibers [32,33]. However, these approaches require 2.5–30 s per point in water, limiting throughput in living specimens. PHBD solves the background problem at the detection level rather than at the fiber level, permitting a standard bare single-mode fiber and reducing acquisition time to 10 ms per point, which represents an improvement of two to three orders of magnitude that brings fiber-coupled Brillouin spectroscopy closer to *in vivo* timescales.

The instrument is modular and predominantly fiber based. Both the pulsed excitation and the CW LO derive from mature telecom-band 1560-nm seed lasers and fiber amplifiers, frequency-doubled in standard periodically poled lithium niobate, so we assembled the system largely from catalog polarization-maintaining components within a compact 62 × 48 × 46 $cm^3$ footprint, the spectrometer occupying 23 × 15 × 2.5 $cm^3$ (Supplementary Fig. 17).

We note several limitations. First, fiber insertion is inherently invasive, and the *in vivo* brain measurements characterize tissue along the insertion track rather than undisturbed tissue. Second, the fiber probe as implemented here provides point spectroscopy; endoscopic lateral mapping through a lensed probe would require distal scanning, which we have not yet demonstrated. Third, the Brillouin shift does not directly measure the longitudinal modulus but also depends on refractive index and mass density (Methods). In many cells and tissues, these quantities co-vary [7,17,26], such that the ratio $\rho/n^2$ varies substantially less than the Brillouin shift, allowing shift variations to be interpreted primarily as changes in longitudinal modulus [17,26,45]. Nevertheless, this assumption should be validated for each new tissue context. Finally, the biological demonstrations presented here are proof of concept; larger cohorts and longitudinal studies will be required to establish the biological significance of the observed regional contrasts.

Several architectural refinements will broaden the scope of PHBD (Supplementary Note 8). A carrier-suppressed electro-optic modulator generating a double-sideband LO would measure the Stokes and anti-Stokes simultaneously, doubling the signal and halving the acquisition time. The 125-μm fiber can be replaced by single-mode fiber as narrow as 80 μm to reduce insertion trauma, and distal beam scanning with miniaturized piezoelectric benders would convert the probe from point spectroscopy to imaging without an external stage. Together, these provide a foundation for endoscopic, catheter- and needle-based Brillouin platforms, and may extend mechanical measurement to sites that benchtop microscopy cannot reach.

## ACKNOWLEDGMENTS

F.Y. acknowledges supports from the Young Scientists Fund of the National Natural Science Foundation of China (23GR1621), the Chinese Academy of Sciences (26XY5409), and the Shanghai Institute of Optics and Fine Mechanics (26J01601). F.Y., L.Z., and Y.Q. acknowledge support from Strategic Priority Research Program of the Chinese Academy of Sciences (XDB0650000).

## AUTHOR CONTRIBUTIONS

F.Y. conceived and initialized the project. Z.D. built the heterodyne Brillouin microscope with help from J.Y. and X.L. F.Y., Z.D., Y.Q., Z.Y. and L.Z. developed the pulse-enhancement theory for heterodyne microscopy. S.Y. built the VIPA Brillouin microscope. Y.Q. built the laser source system with help from W.C. J.Z. and Z.D. developed the acquisition and analysis code. Z.D. performed the biological experiments and analyzed the data, with help from Y. Liu, Y. Luo, T.M., Y.W., G.Z., F.H., J.X., S.C. and W.Y. F.Y., W.C., L.Z. and Y.Q. led the project. F.Y., Z.D. and Y.Q. wrote the paper with input from all authors.

## COMPETING INTERESTS

F.Y., Z.D. and Y.Q. are inventors of a patent related to the technology described in this paper. All other authors declare no competing interests.

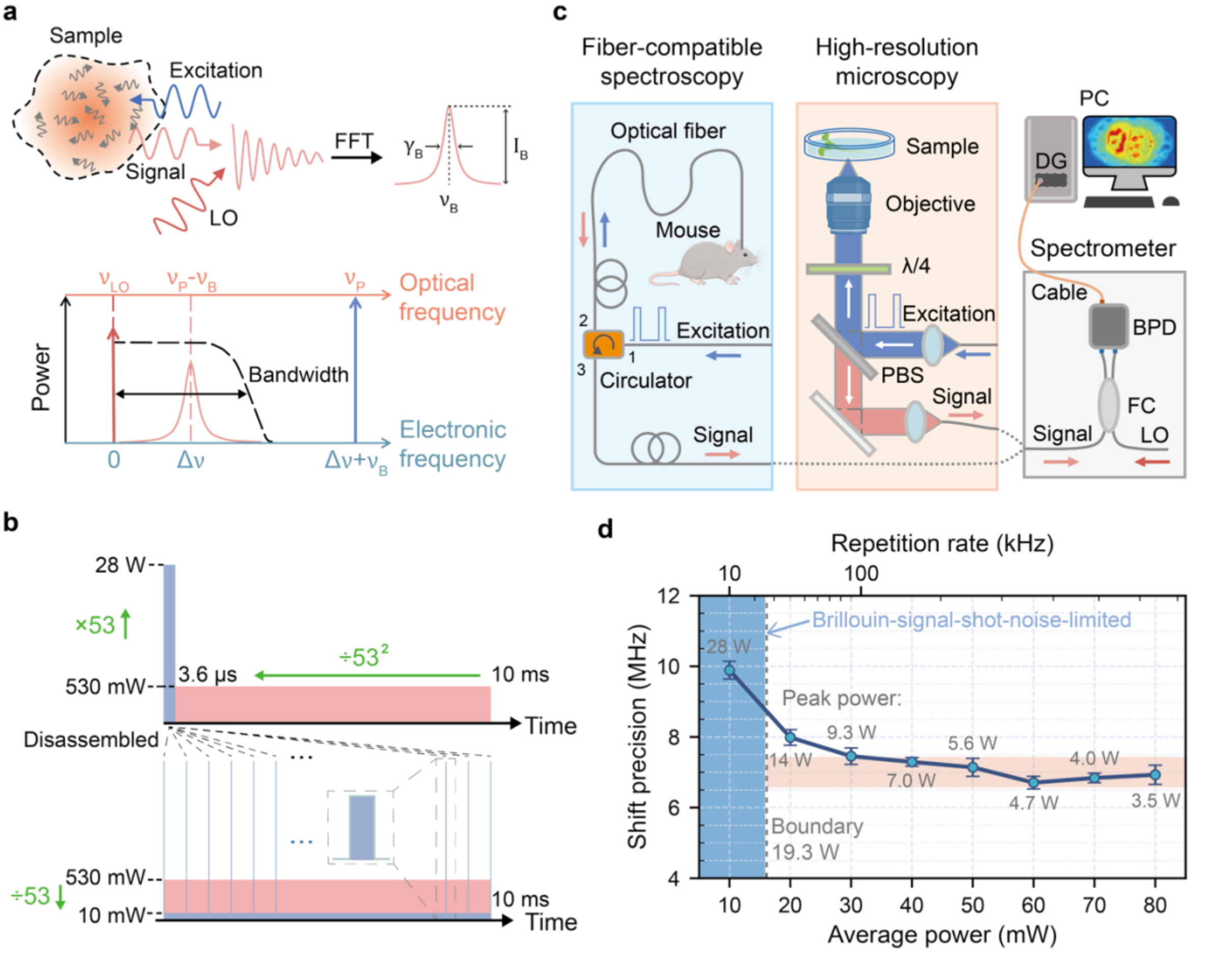

**Fig. 1 Principle and implementation of pulsed heterodyne Brillouin detection. a**, Principle of heterodyne Brillouin detection. Excitation light interacts with thermally excited acoustic phonons in the sample to generate frequency-shifted Brillouin scattering. Mixing the backscattered Brillouin signal with a local oscillator (LO) produces a temporal beat note, whose Fourier transform yields the Brillouin spectrum and hence the shift $\nu_B$, the linewidth $\gamma_B$ and the intensity $I_B$ (upper panel). Heterodyne mixing maps optical frequencies onto electronically detectable beat frequencies referenced to the LO; only components within the electronic detection bandwidth are recorded (lower panel). **b,** Pulse-enhancement principle at a fixed signal-to-noise ratio (SNR). Starting from the CW reference condition, raising the excitation power 53-fold, from 530 mW to 28 W, shortens the effective excitation duration needed to maintain precision under 10 MHz by a factor of $53^2$ (~2,800), from 10 ms to 3.6 µs, and thereby reduces the delivered optical energy 53-fold (upper panel). Redistributing the 3.6 µs of excitation as a pulse train across a 10-ms acquisition window maintains a peak power of 28 W while lowering the average power to 10 mW, 53-fold below the CW reference (lower panel). **c**, Optical layouts of the fiber-compatible spectroscopy and high-resolution microscopy configurations, which share a common heterodyne spectrometer. In the fiber-compatible configuration, pulsed excitation is routed through an optical circulator and delivered through an optical fiber that also collects the backscattered Brillouin signal. In the microscopy configuration, excitation is focused onto the sample through an objective, and the backscattered signal is separated from the excitation path by a quarter-wave plate (λ/4) and polarizing beam splitter (PBS). In both, the Brillouin signal is combined with the LO in a fiber coupler (FC), detected by a balanced photodetector (BPD) and digitized for spectral reconstruction. DG, digitizer; PC, personal computer. **d**, Brillouin shift precision measured in water in the microscopy configuration, as a function of average excitation power at a fixed 10-ms

acquisition time. We held the pulse width at 36 ns and reduced the repetition rate in proportion to the square of the average power, resulting in an inverse relationship between the average and peak powers. The corresponding peak powers are indicated next to the data points, and the pulse repetition rates are shown on the upper axis. Data are mean ± s.d. of five independent measurements, the precision of each estimated from $n$ = 400 spectra. The gray dashed line marks the transition between the constant-noise-limited and Brillouin-signal-shot-noise-limited regimes. In the orange region (constant-noise-limited), the shift precision remains invariant as peak power rises and average power falls, as predicted; in the blue-gray region (Brillouin-signal-shot-noise-limited), precision degrades as the average power decreases at fixed acquisition time.

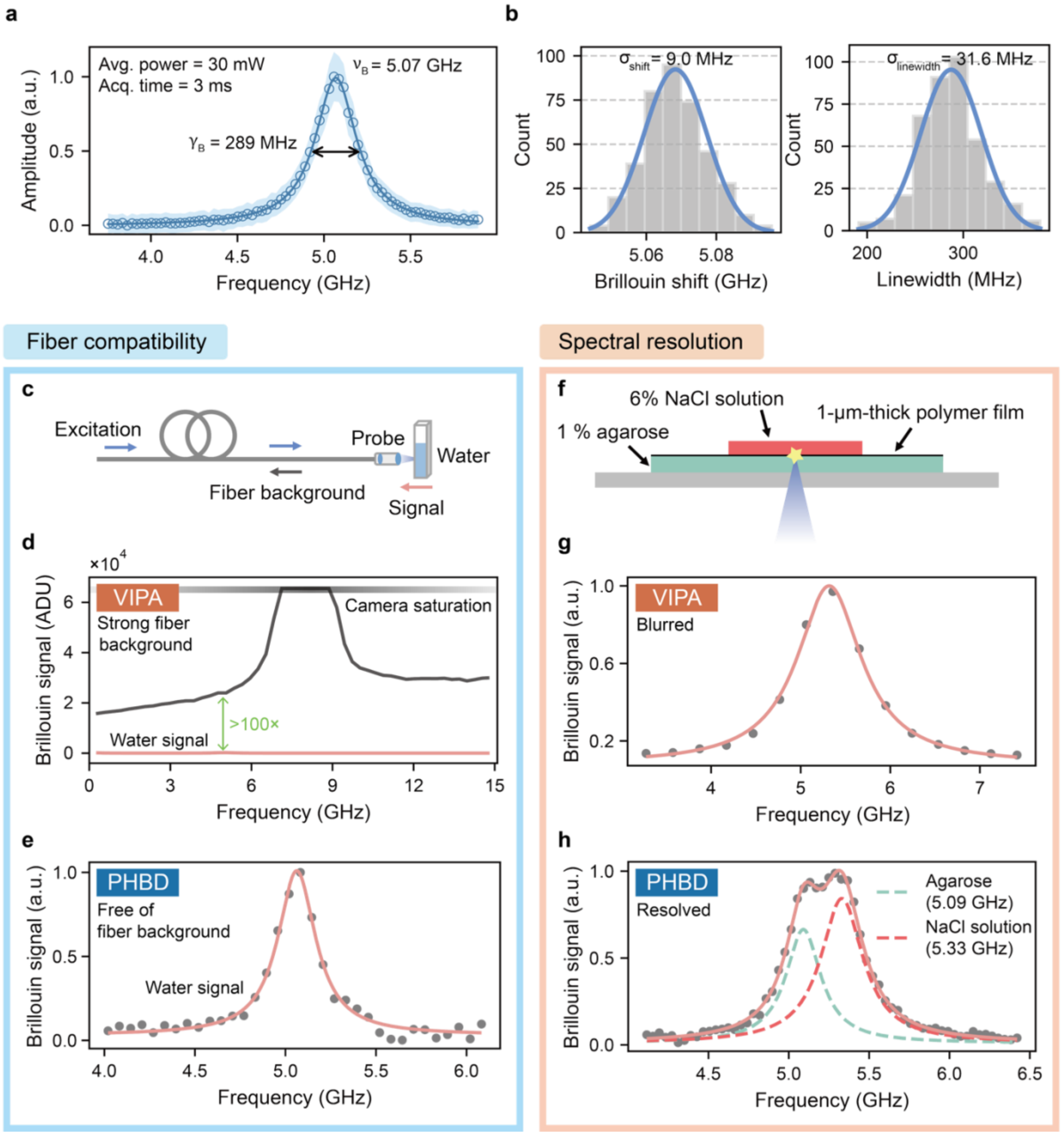

**Fig. 2 Precision, fiber-background rejection and spectrometer resolution of PHBD. a**, Representative Brillouin spectrum of water acquired through a low-NA objective (NA=0.12) at 30-mW average excitation power and a 3-ms acquisition time. Open circles show the mean of $n$ = 400 spectra, light-blue shading ±1 s.d. and the solid blue curve the Lorentzian fit. **b**, Distributions of the Brillouin shifts and linewidths fitted to the $n$ = 400 individual spectra of **a**. Gray histograms show the measured distributions and blue curves the Gaussian fits; the standard deviations of the fitted shift and linewidth were 9.0 MHz and 31.6 MHz, respectively. **c,** Schematic of the fiber-coupled configuration used to measure the Brillouin signal of water. We used pulsed excitation for PHBD and continuous-wave excitation for the VIPA spectrometer, through the same 0.5-m delivery fiber. **d,** Intensity profile recorded on the sCMOS camera during a fiber-coupled water measurement with the VIPA spectrometer. At 45-mW excitation power and a 50-ms exposure, the strong background generated within the delivery fiber saturated the camera and exceeded the water signal by at least 100-fold within the spectral region of interest. **e,** Representative Brillouin spectrum of water acquired with fiber-compatible PHBD through the same fiber at 20-mW average power and a 10-ms

acquisition time, giving a Brillouin shift precision of 8.9 MHz ($n$ = 400 spectra). **f,** Schematic of the interface between 6% (w/v) NaCl solution and 1% (w/v) agarose. A 1-μm-thick polymer film prevented diffusion of the NaCl solution into the agarose. **g,h,** Representative Brillouin spectra acquired at the NaCl-solution–agarose interface with the VIPA spectrometer (**g**) and PHBD (**h**), using the same objective at an effective NA of 0.24. The VIPA spectrometer returned a single broadened peak, whereas PHBD resolved two components, at 5.09 GHz (agarose) and 5.33 GHz (NaCl solution), which we quantified by double-Lorentzian fitting.

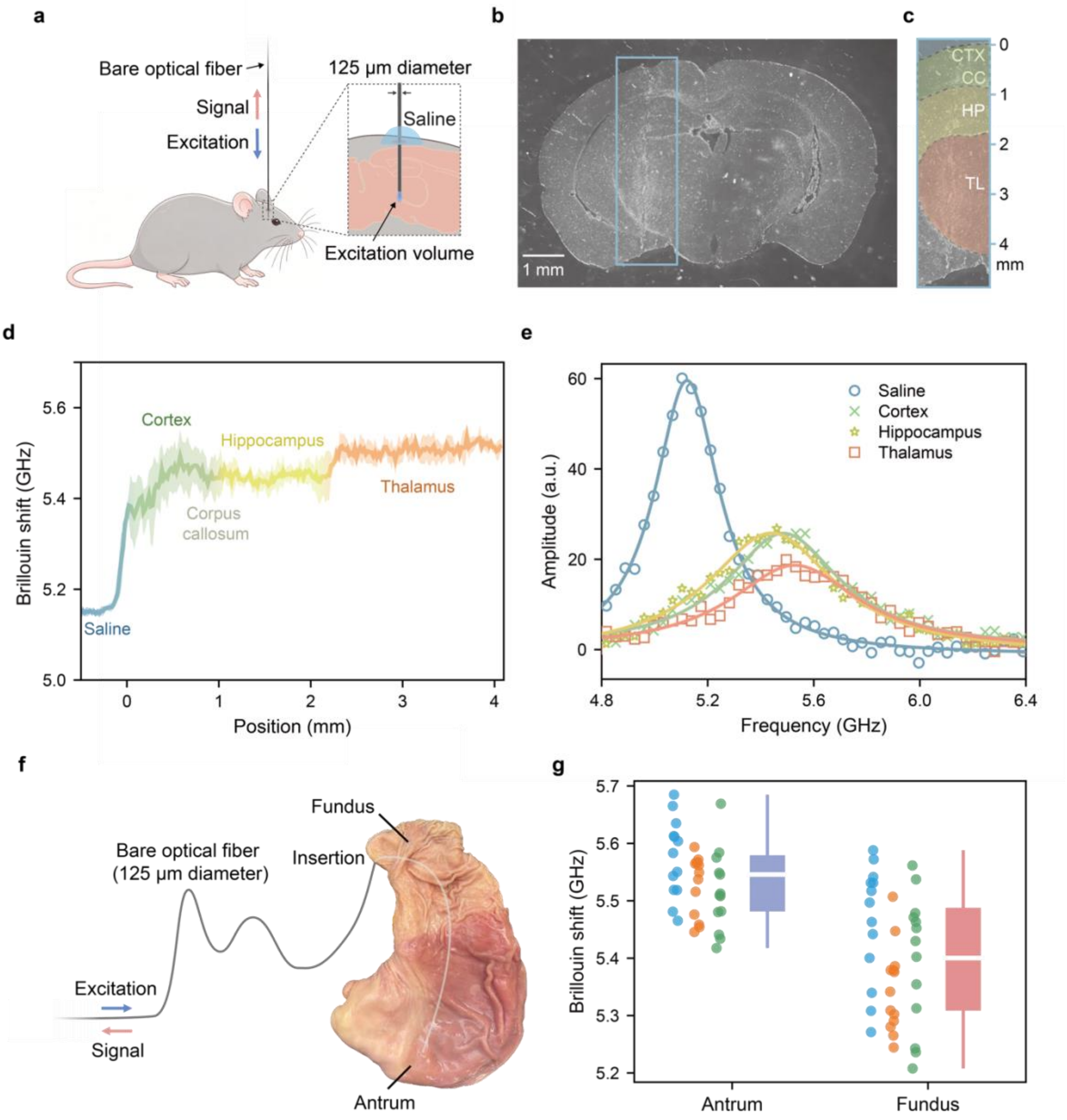

**Fig. 3 Fiber-coupled Brillouin spectroscopy in the mouse brain *in vivo* and in an intact porcine stomach. a,** Schematic of the mouse brain experiment. A 125-µm-diameter bare optical fiber both delivered excitation light to deep brain and guided the backscattered Brillouin signal to the detector. We mounted the mouse on a *z*-axis translation stage while the fiber remained fixed above the cranial opening, so that raising the stage advanced the fiber progressively deeper. Saline applied at the insertion site prevented tissue dehydration. **b,** Dark-field image of a paraffin-embedded brain section after fiber insertion, with the insertion track outlined in blue. **c,** Enlarged view of the boxed region in **b**, showing the anatomical regions traversed by the fiber: cortex (CTX), corpus callosum (CC), hippocampus (HP) and thalamus (TL). Owing to its limited thickness, the corpus callosum is not delineated separately. **d**, Depth-resolved Brillouin-shift profile measured along the insertion track in 20-µm steps at 25-mW average power ($n = 3$ mice). The solid line indicates the mean, the shaded band ±1 s.d., and the colors the anatomical regions identified in **c**. We set the acquisition time to 50 ms per depth to raise the SNR in tissue. **e,** Representative Brillouin spectra acquired from saline, cortex, hippocampus and thalamus along the insertion track. **f**, Schematic of luminal Brillouin spectroscopy in a freshly excised intact porcine stomach. We guided a bare optical fiber through the stomach lumen to deliver excitation light and collect Brillouin signals from the antrum and fundus. **g**, Brillouin shifts measured at spatially distributed sites in the antrum (39 sites from $n = 3$ biologically independent stomachs) and fundus (39 sites from $n = 3$ biologically independent stomachs). Each point represents one measurement site, and colors denote individual stomachs. Boxes indicate the interquartile range, white center lines indicate the median, and whiskers extend to the most extreme values within 1.5× the interquartile range.

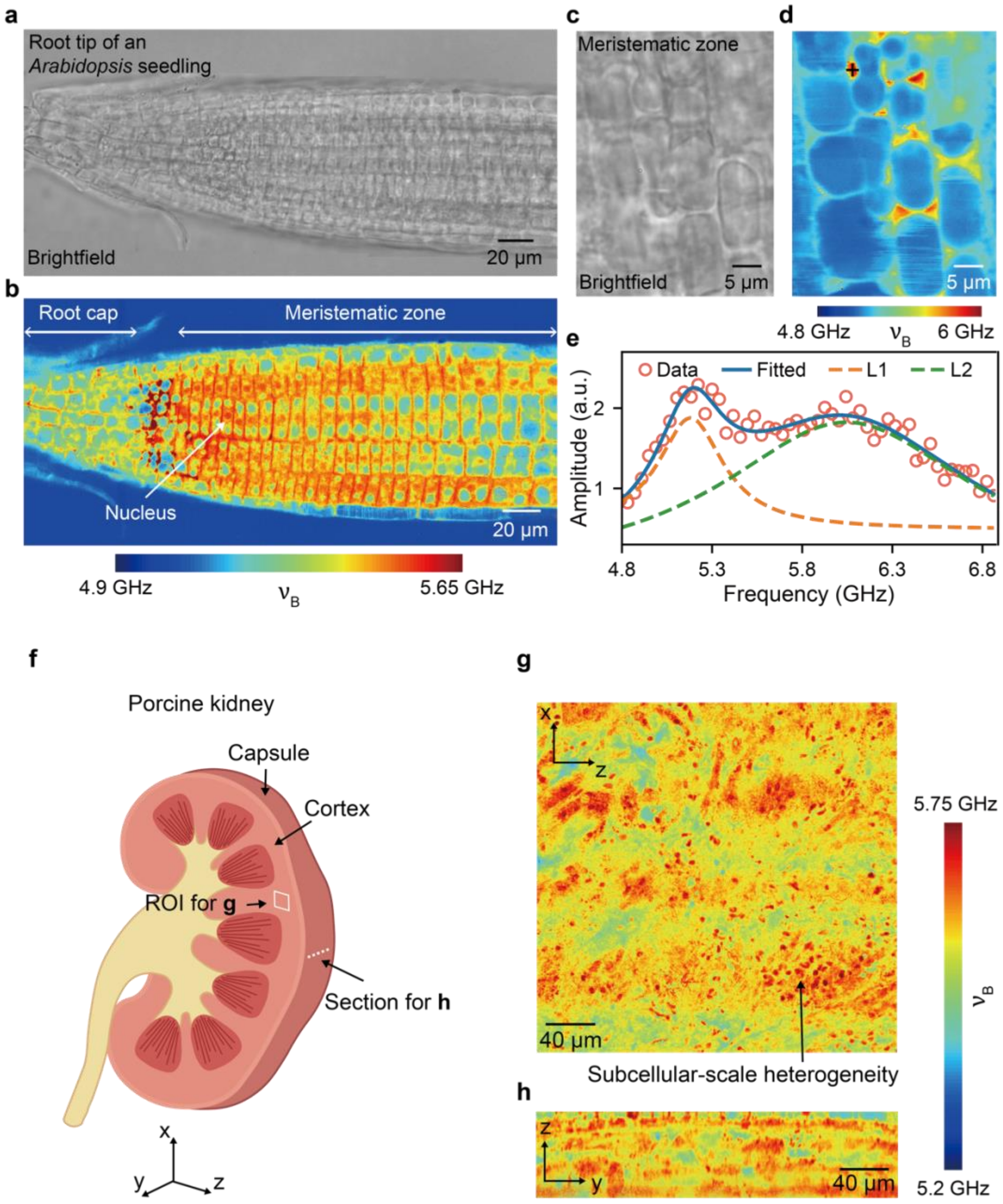

**Fig. 4 High-specificity Brillouin imaging of *Arabidopsis* root tips and porcine renal cortex.** **a**,**b**, Brightfield image of the root tip of an *Arabidopsis* seedling (**a**) and the corresponding Brillouin shift map (**b**). Scan area, 130 µm × 280 µm; step, 0.25 µm × 0.25 µm. For visualization only, we assembled the brightfield image in **a** from ten partially overlapping fields of view; we filled blank edge regions introduced by stitching with a background image; the Brillouin shift map in **b** was not modified. **c,d,** Brightfield image of the meristematic zone of an *Arabidopsis* root tip (**c**) and the corresponding Brillouin shift map (**d**). Scan area, 35 µm × 45 µm; step, 0.2 µm × 0.2 µm. **e,** Representative two-peak spectra acquired at the cell wall-cytoplasm interface, resolving components at 5.18 GHz (cytoplasm) and 6.05 GHz (cell wall). **f,** Schematic of a porcine kidney cross-section, indicating the region of interest (ROI) for **g** and the section plane for **h.** **g**, Brillouin shift image of the *x*–*z* plane within the ROI in **f**, revealing subcellular-scale mechanical heterogeneity. Scan area, 300 µm × 280 µm; step, 0.4 µm × 0.4 µm. **h**, Cross-sectional Brillouin shift map of the *y-z* plane along the dashed line in **f**. Scan area, 300 µm × 70 µm; step, 0.2 µm × 0.2 µm. We acquired all images using a high-NA objective (NA = 0.7).

## Methods

### Brillouin light scattering

Spontaneous Brillouin scattering originates from the inelastic interaction of incident photons with thermally driven acoustic phonons in the probed material. A photon that transfers energy to a phonon is downshifted in frequency (Stokes component), whereas a photon that gains energy from a thermally populated phonon is upshifted (anti-Stokes component). For an isotropic medium, conservation of energy and momentum sets the phase-matching condition $\nu_B = 2n/\lambda \sqrt{M'/\rho}\,\sin(\theta/2)$, where $\nu_B$ is the Brillouin frequency shift, $n$ the refractive index, $\lambda$ the vacuum wavelength of the incident light, $\rho$ the local mass density, $M'$ the longitudinal modulus and $\theta$ the angle between the incident and scattered wavevectors. In the backward (epi-) detection geometry ($\theta = \pi$), this reduces to $M' = \rho\lambda^2\nu_B^2/(4n^2)$.

### Laser source system

Both the pulsed excitation laser and the CW LO were built on a polarization-maintaining master-oscillator power-amplifier (MOPA) architecture followed by second-harmonic generation (SHG), shown schematically in Supplementary Fig. 4a and in detail in Supplementary Fig. 4b.

The short-wave infrared (SWIR) excitation seed was an external-cavity diode laser (ECDL) centered at 1560.49 nm. It was chopped and shaped into seed pulses by a semiconductor optical amplifier and an acousto-optic modulator, giving nanosecond pulse widths at repetition rates of tens of kilohertz. The seed pulse train was pre-amplified in Er-doped fiber (EDF; PM-ESF-7/125, Coherent) and then boost-amplified in Er/Yb co-doped double-cladding fiber (EYDF; PM-EYDF-10/125-XPH, Coherent). At a pulse width of 25.6 ns and a repetition rate of 30 kHz, the quasi-rectangular SWIR pulses reached a peak power of 350 W. The SWIR pulses were collimated and focused by a plano-convex lens into a periodically poled lithium niobate crystal (BC-PPLN; MSHG1550-1.0-40, Covesion) with an SHG efficiency of ~70%. The frequency-doubled beam passed a dichroic mirror that removed the residual fundamental beam, then double-passed an etalon (OP-7423-6743-2, LightMachinery) to suppress amplified spontaneous emission. A typical 780-nm pulse delivered to the Brillouin microscope is shown in Supplementary Fig. 4c.

The LO was seeded by a second SWIR ECDL, detuned ~3.5 GHz below the excitation seed. After amplification in another piece of EDF, the LO seed was frequency-doubled in a waveguide PPLN (WG-PPLN; Honclabs), giving 45 mW of output at a conversion efficiency of 12%. The CW LO was then delivered to the fiber coupler and mixed with the Brillouin signal for heterodyne detection. A photograph of the complete laser system, mounted on a movable cart, is shown in Supplementary Fig. 17b. Because the laser system was under continuous development throughout this work, pulsed-laser parameters differed between experiments; the parameters used for each are listed in Supplementary Table 2.

### Fiber-compatible spectroscopy and free-space imaging configurations

A detailed schematic of the fiber-compatible spectrometer is shown in Supplementary Fig. 5a. Two probe configurations were used: a lensed probe and a bare-fiber probe. Pulsed excitation entered port 1 of an optical circulator and was directed from port 2 to the probe. The same probe collected the backscattered Brillouin signal, which returned through port 2 and was routed from port 3 to the heterodyne spectrometer. The lensed probe comprised a pair of aspheric lenses (Supplementary Fig. 5b): the first (AC91536, LBTEK) collimated the light emerging from the fiber and the second (OLSM041509-T3, JCOPTIX) focused it onto the sample. In the bare-fiber configuration, the fiber delivered excitation directly to its distal end and collected the backscattered signal; the distal facet was angle-cleaved to suppress back-reflection at the glass–air interface.

A complete schematic of the free-space configuration is shown in Supplementary Fig. 6 and a photograph of the setup in Supplementary Fig. 17c. Pulsed excitation was delivered to the microscope through a PM780-HP fiber and collimated by a fiber collimator (AFC1550-2.1-APC, LBTEK). The beam polarization was set to linear p-polarization with a quarter-wave plate and a half-wave plate. The p-polarized beam was transmitted through a polarizing beam splitter (PBS) and converted to left-circular polarization by a second quarter-wave plate. A beam expander (OSAE02-T3, JCOPTIX) increased the beam diameter to 4.4 mm, filling the back aperture of the objective (LUCPLFLN60X, Olympus) so that its full numerical aperture was used. Backscattered spontaneous Brillouin light was collected by the same objective and propagated back through the beam expander and quarter-wave plate, which converted it to s-polarization. The s-polarized signal was reflected by the PBS and coupled into the signal fiber through an identical fiber collimator.

A photograph of the PHBD spectrometer is shown in Supplementary Fig. 17d. The signal and LO fibers were connected to the two input ports of a 2 × 2, 50:50 polarization-maintaining fiber coupler, whose output ports were directed to a balanced photodetector (BPD). The resulting electrical signal was sampled by a digitizer installed directly in the computer, giving a compact spectrometer assembly with dimensions of 23 × 15 × 2.5 $cm^3$. Two detection-bandwidth configurations were used: a 1.6-GHz BPD paired with a 1.25-GHz digitizer, giving an effective detection bandwidth of 1.25 GHz; and a 2.5-GHz BPD paired with a 2.5-GHz digitizer, giving an effective detection bandwidth of 2.5 GHz.

**Data processing and spectral calibration**

To retrieve the Brillouin spectrum from the time-domain signal, a fast Fourier transform (FFT) was applied to each data segment, whose duration was set by the pulse width (Supplementary Note 1). Because the Brillouin spectrum corresponds to the power spectrum of the heterodyne signal, we evaluated the normalized squared magnitude of the Fourier-domain signal. The resulting spectrum contains a detection-system-dependent baseline arising predominantly from LO shot noise (Supplementary Fig. 2). To remove it, a high-SNR background spectrum was obtained by averaging spectra acquired over a long interval, typically 4 s, in the absence of the Brillouin signal, and was subtracted from each measured spectrum. Because the variance of the averaged background was substantially lower than that of an individual measurement, the subtraction degraded the overall SNR negligibly. The background-subtracted spectra acquired during each measurement were then averaged, the number averaged being set by the acquisition time and the pulse repetition rate, and the averaged spectrum was normalized by the measured frequency-dependent responsivity of the PHBD spectrometer to compensate for its non-flat spectral response (Supplementary Fig. 8).

The beating frequency ($f_{LP}$) between the LO and the excitation seed lasers was recorded before experiments. Because the acousto-optic modulator introduced a frequency shift of 300 MHz, the beat frequency between the LO and the pulsed excitation light was calculated as $f'_{LP} = f_{LP} + 300\ MHz$. Because PHBD detects the Stokes component, with the LO frequency lower than the Brillouin-signal frequency, the frequency axis $x_f$ was therefore converted to $x'_f = 2 \times f'_{LP} - x_f$ to recover the corresponding Brillouin frequency axis. This procedure provides direct frequency referencing and avoids the more involved spectral calibration required for VIPA-based spectrometers, reducing calibration uncertainty.

## Performance characterization

### Measurement of spectral resolution

Spectral resolution was first evaluated from the measured spectra of the pulsed laser (Supplementary Fig. 10a–c). The objective was replaced by a high-reflectivity mirror, all other components remaining unchanged, and the excitation frequency was detuned from the Rb cell absorption band so that the beam entered the spectrometer without absorption. The LO frequency was adjusted to maintain a fixed beat frequency of 420 MHz with the excitation laser. Pulsed-laser linewidth and FFT processing were assessed independently as sources of spectral broadening. The pulse width was fixed at 204.8 ns while the FFT window duration was varied (Supplementary Fig. 10a). When the pulse width was varied, the FFT window was fixed at 25.6 ns. For example, the 128-point time-domain data acquired from a 51.2-ns pulse was divided into two 64-point segments, each corresponding to a 25.6-ns window (digitizer bandwidth, 1.25 GHz; Supplementary Fig. 10b). Excluding NA-induced broadening, the dominant contribution came from the finite temporal window used for FFT analysis. When the number of sampling points was limited and the measured laser spectrum was very narrow, symmetric zero-padding was applied before and after the acquired time-domain data to extend the total length to 512 points. This operation interpolated the frequency grid and facilitated more stable spectral fitting, without changing the intrinsic spectral resolution (Supplementary Fig. 10c).

The linewidth of water was then measured at several pulse widths through a low-NA (0.12) objective as an independent check (Supplementary Fig. 10d). After subtracting a 263.7 MHz offset, the measured trend followed an inverse-proportional function, indicating a total spectral broadening of 35.0 MHz at a pulse width of 25.6 ns, in agreement with the directly measured laser spectral width of 37.7 MHz (Supplementary Fig. 10a).

The spectral resolution at the 36-ns pulse width used for imaging was estimated by subtracting the intrinsic linewidth of water (263.7 MHz) from the measured linewidth (289 MHz; Fig. 2a), yielding an instrumental linewidth of approximately 25 MHz, in close agreement with the theoretical estimate of 27 MHz (Supplementary Fig. 10a,b). NA-induced broadening was neglected owing to the low effective NA of 0.12.

### Comparison of PHBD and stimulated Brillouin microscopy

To confirm the accuracy of the Brillouin spectra obtained with PHBD, the same water sample was measured at the same room temperature with the free-space PHBD configuration and with an SBS microscope configured as in ref. 40 (Supplementary Fig. 11). To minimize the pulse broadening, a 60-ns pulse width was used in the SBS measurement, contributing approximately 2 MHz of additional spectral broadening (Supplementary Fig. 10b). To eliminate broadening introduced by the lock-in amplifier, we used a very slow frequency-scanning rate (150 ms at 0.15 NA and 100 ms at 0.7 NA) together with a high noise-equivalent power bandwidth of 800 Hz. For PHBD, a 36-ns pulse width was used throughout, giving a theoretical spectral broadening of 27 MHz, comprising 4 MHz from the pulsed-laser linewidth and 23 MHz from FFT processing (Supplementary Fig. 10a,b). After these contributions were subtracted, the residual water linewidths reflect only NA-induced broadening and agree closely between the two systems.

### Measurement of spatial resolution

For fiber-compatible configuration with a lensed probe, lateral resolution was evaluated by imaging group 7 of a positive 1951 USAF resolution target with the transmitted laser power recorded by a photodetector (Supplementary Fig. 5f). Edge profiles across the target were fitted with error functions, and the FWHM of the derivative of each fitted curve gave lateral resolutions of 1.8 μm and 1.8 μm along the two transverse directions (Supplementary Fig. 5g,h). Axial resolution was evaluated by imaging the interface between double-distilled water and a coverslip. Only the water signal was detected, because the Brillouin shift of borosilicate glass (>20 GHz) lies far outside the detection bandwidth. The axial amplitude profile was fitted with an error function, yielding an axial resolution of 17.4 μm (Supplementary Fig. 5i–k).

For the free-space configuration, lateral resolution was evaluated by imaging the central plane of a polymethyl methacrylate (PMMA) bead embedded in 3% (w/v) agarose. The Brillouin shift of PMMA is approximately 10.8 GHz, estimated from its refractive index of 1.49 and acoustic velocity of 2820 m/s, far above that of agarose (~5.02 GHz). Therefore, within the detection bandwidth, only the agarose signal was detected, avoiding spectral crosstalk from the bead. Brillouin signal amplitudes obtained by fitting the measured spectra were used to generate the amplitude map (Supplementary Fig. 7b). Edge profiles across the agarose-PMMA boundary were fitted with error functions along x and y, and the FWHM of the derivative of each fitted curve gave lateral resolutions of 0.6 μm and 0.6 μm (Supplementary Fig. 7c,d). Axial resolution was evaluated by the water–coverslip method described above, giving 2.4 μm (Supplementary Fig. 7g).

**Determination of spectral precision**

Double-distilled water was used to evaluate Brillouin-shift precision. The Brillouin spectrum of water was acquired repeatedly and the precision quantified as the standard deviation of the Lorentzian-fitted center frequencies across $n = 400$ spectra. Brillouin shift precision at different combinations of average and peak excitation powers (Fig. 1d) was measured at a fixed 10-ms acquisition time using 36-ns pulses and a low-NA objective (ACHN10XP; effective NA, 0.12). The beat frequency between the excitation laser and the LO was set to 5.58 GHz, corresponding to a detected spectral range of 4.33–5.58 GHz. The product of the average and peak powers was held constant by adjusting the pulse repetition rate in proportion to the square of the average power.

**Sample preparation and viability test**

**Chemical samples.** The water sample was prepared by dropping 2 mL double-distilled water in a glass-bottomed Petri dish (MatTek, P35G-1.5-14-C). The NaCl solution-agarose sample was made by first dropping 30 μL 1% agarose on the surface of a #1.5 coverslip. Before the gelling of agarose, a 1 cm × 1 cm size polymer thin film with 1 μm thickness was attached to the agarose, followed by immediate addition of 30 μL 6% w/v NaCl solution on the top flat area. To prepare the PMMA beads in agarose, the PMMA bead was first diluted to 0.1% w/v. Then 8 μL 5% agarose was dropped on the middle of a #1.5 coverslip, on which a 120-μm-thick spacer (Grace Bio-Labs SecureSeal) was attached beforehand. Five microliters of the 0.1% (w/v) PMMA-bead suspension was immediately added into the agarose drop, then another coverslip was attached by the imaging spacer for sealing the sample.

**Mice preparation.** All mouse experimental procedures were approved (AD2026131) by the Animal Ethics Committee of Shanghai Songjiang District Chedun Experimental Animal Breeding Farm Co., Ltd. Three male wild-type C57BL/6J mice, aged 22–24 weeks at the time of surgery, were used. Acute, non-recovery cranial surgery was performed under isoflurane anesthesia. Mice were anesthetized with 2% isoflurane vaporized in medical-grade $O_2$ in an induction chamber, then placed prone in a stereotaxic frame and maintained under anesthesia by nose-cone delivery of 1.5–2% isoflurane using a small-animal gas-anesthesia system. A surgical plane of anesthesia was confirmed by the absence of the pedal-withdrawal reflex before incision. Anesthesia depth was monitored by respiratory rate and reflex responses, and the isoflurane concentration was adjusted as needed. A circular portion of skull (400 μm in diameter) over the left primary somatosensory cortex (relative to bregma: AP, −2.06 mm; ML, −1.50 mm) was removed with a dental drill under constant phosphate-buffered saline (PBS) irrigation, leaving the dura intact. At the end of the experiment, mice were deeply anesthetized, and euthanasia was performed by cervical dislocation. Death was confirmed by cessation of respiration and heartbeat.

**Porcine stomach and kidney.** Fresh porcine stomach and kidney tissues were purchased from a local supermarket. No live animals were used or euthanized for this study. Because the tissues were commercially obtained as food-grade products, institutional animal ethics approval was not required. For spectroscopic measurements, the stomach was used intact. For imaging, the stomach and kidney tissues were cut into small pieces, placed in Petri dishes and immersed in PBS to prevent dehydration during measurement.

**Root tip of *Arabidopsis* seedling.** *Arabidopsis thaliana* seeds were sterilized in 70% ethanol containing 0.05% Triton X-100 for 5 min, washed in 95% ethanol for 1 min, air-dried, and sown on half-strength Murashige-Skoog medium supplemented with 0.05% MES and 0.8% Phytoagar (pH 5.7). After stratification at 4 °C for 2–3 days, seeds were germinated under long-day conditions, with a 16-hour light/8-hour dark photoperiod, a light intensity of 170 μmol/(m$^2$·s), and day/night temperatures of 21 °C/17 °C. Seedlings with clearly developed primary roots were selected for imaging. Before imaging, 100 μL of water was placed on a #1.5 coverslip, and an individual seedling was transferred into the water droplet. The seedling was gently covered with a small piece of solidified Phytagel pad to reduce dehydration during imaging.

**Cells.** Cell lines were obtained from Cobioer and maintained according to the guidelines provided by the American Type Culture Collection. Mouse fibroblasts (NIH/3T3, CBP60317) and human cervical carcinoma cells (HeLa, CBP60232) were cultured in Dulbecco's modified Eagle's medium (DMEM; Gibco, 11960044) supplemented with 10% (v/v) fetal bovine serum (Gibco, A5669701) and 100 U mL$^{-1}$ penicillin–streptomycin (Gibco, 15070063). Cells were seeded onto glass-bottom dishes (Titan confocal Petri dish, BDD012035) or polyacrylamide-gel-coated dishes (Matrigen, SV3510-EC-12) at a density of 5,000 cells cm$^{-2}$ and allowed to adhere overnight before imaging.

***C. elegans* embryos.** *C. elegans* adults of the wild-type N2 Bristol strain were rinsed from an agar plate with 1 mL of M9 buffer (22 mM $KH_2PO_4$, 49 mM $Na_2HPO_4$, 86 mM NaCl and 10 mM $NH_4Cl$) and transferred into a 1.5 mL centrifuge tube. After centrifugation, the supernatant was removed, and 1 mL of worm bleach solution (a 1:1 mixture of 10% NaClO and 1 M NaOH) was added. The sample was gently agitated for 2-3 min to lyse the adult bodies and release the embryos. Following a second centrifugation, the supernatant was removed and replaced with 1 mL of M9 buffer. The mixture was vortexed to wash the embryos and centrifuged again. The washing step was repeated twice. For mounting, 15 μL of 1% agarose was deposited on a coverslip, onto which two imaging spacers (combined thickness: 240 μm) were stacked and adhered. Immediately afterward, 8 μL of the embryo-containing M9 buffer at the bottom of the centrifuge tube was pipetted onto the agarose drop, and a second coverslip was placed on top to seal the sample.

**Cell viability test.** To assess cell viability under the free-space PHBD configuration, propidium iodide (PI; Sigma-Aldrich, P4170-10MG) was added to the fibroblast culture dishes at 30 μL PI solution per 2 mL of medium. Several regions of interest (ROIs) were selected and marked using the brightfield modality for subsequent imaging. Wide-field fluorescence images were first acquired using the SBS microscope setup, with excitation (MF525-39, Thorlabs) and emission (MF620-52, Thorlabs) filters placed after the light-emitting diode and before the camera, respectively, to confirm cell viability prior to Brillouin imaging. The dishes were then transferred to the free-space PHBD setup for imaging of all cells within the ROIs. After Brillouin imaging, the dishes were returned to the fluorescence microscope to verify cell viability again.

**Data and Code Availability**

The data and code used to generate the results and figures presented in this study will be deposited in Zenodo and will be made publicly available upon publication of the manuscript.

# SUPPLEMENTARY INFORMATION

# Pulsed heterodyne detection enables fiber-compatible, high-specificity Brillouin biomechanics in intact tissue and the living brain

## Table of contents

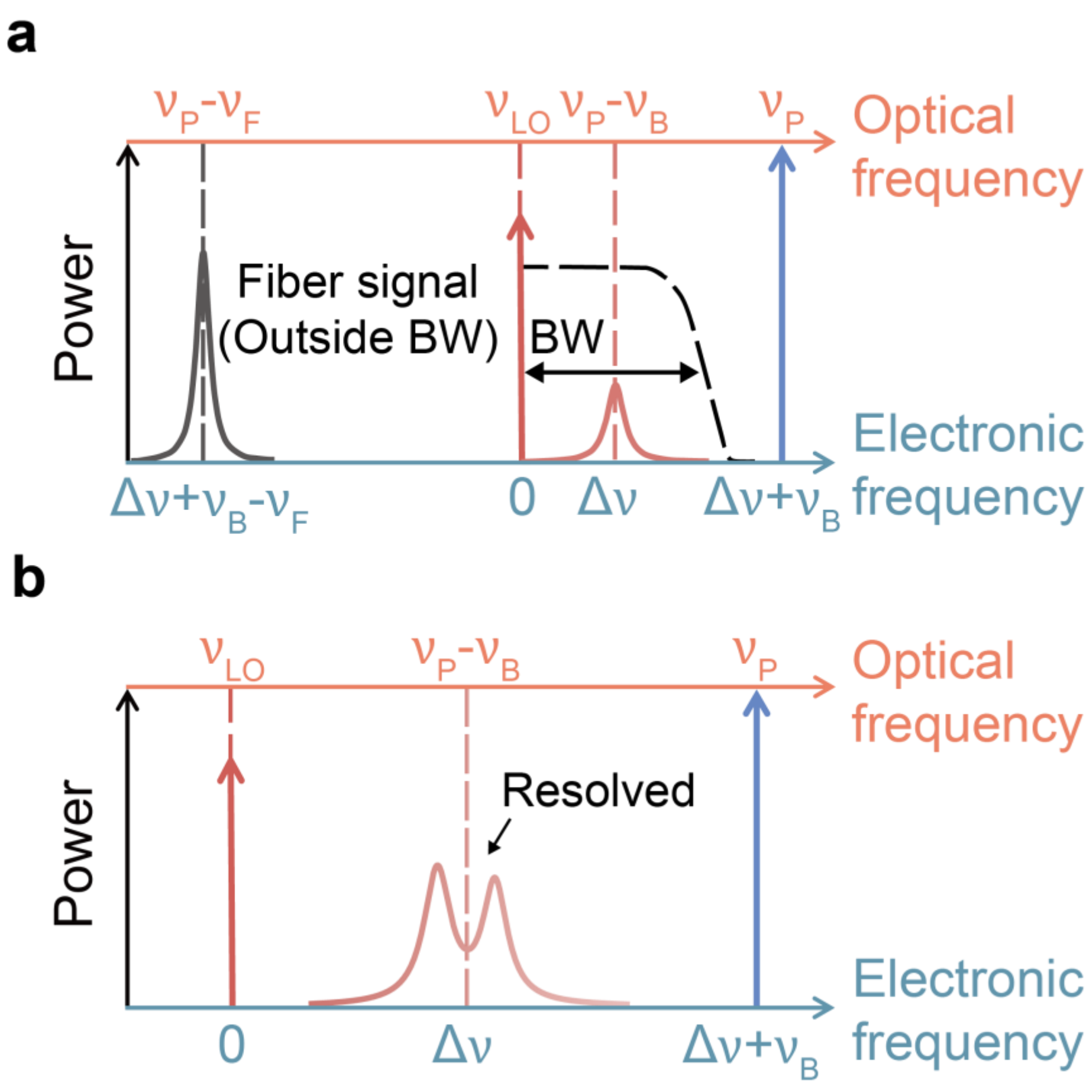

**Supplementary Fig. 1 Frequency-domain characteristics of heterodyne detection.** **a,** The Brillouin signal falls within the electronic detection bandwidth (BW), whereas the strong fiber-generated signal falls outside the bandwidth and is therefore intrinsically rejected. **b,** Mechanically similar biological components within a single focal volume can generate closely spaced Brillouin peaks, which are resolved owing to the narrow instrumental response of heterodyne detection.

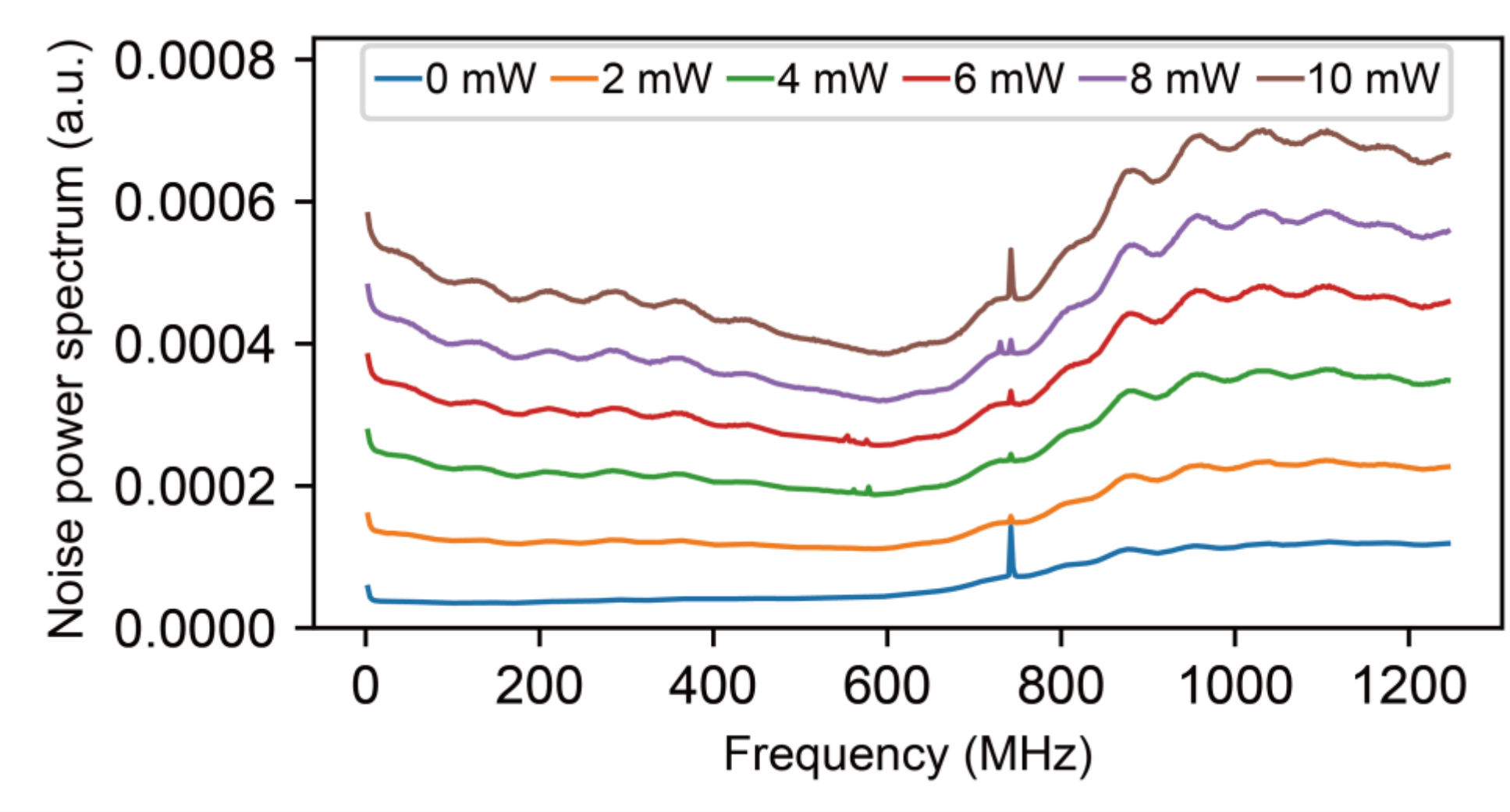

**Supplementary Fig. 2 Characterization of the constant, signal-independent noise floor.** Noise power spectrum of the detection system as a function of frequency for different local oscillator (LO) powers. The trace labeled 0 mW corresponds to the electronic noise floor of the detection system. As the LO power increases, the noise power spectrum scales linearly due to the contribution of optical shot noise, which becomes dominant above 8 mW. In all imaging experiments, the LO power was fixed at 9 mW to ensure LO shot-noise dominance over the electronic noise floor.

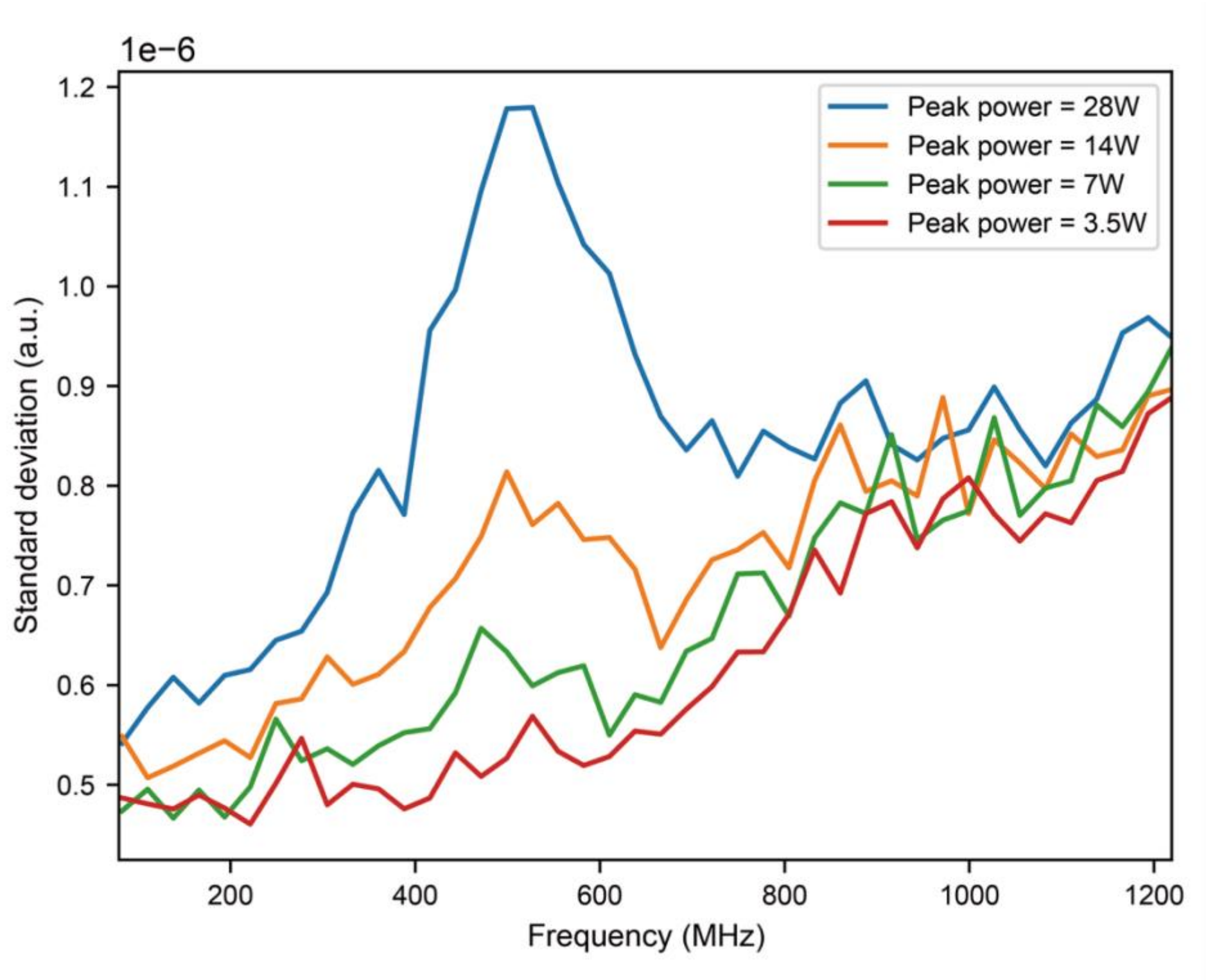

**Supplementary Fig. 3 Dependence of spectral standard deviation on excitation peak power.** The standard deviation of the measured water spectra is plotted as a function of frequency for different laser peak powers. Each curve was calculated from 400 individual raw spectra acquired with the same pulse width of 36 ns, repetition rate of 10 kHz and exposure time of 10 ms. At a peak power of 3.5 W, the Brillouin-signal shot noise contribution is negligible, and the measured standard deviation therefore reflects the baseline noise floor dominated by local oscillator shot noise. The non-flat spectral shape of this noise floor arises from the frequency response of the balanced detector and digitizer, together with the detector's non-white noise spectrum. As the peak power increases, the standard deviation increasingly follows the Brillouin spectral profile, indicating that Brillouin-signal shot noise becomes dominant at a peak power of 28 W. For the 28-W condition, the average power was 10 mW.

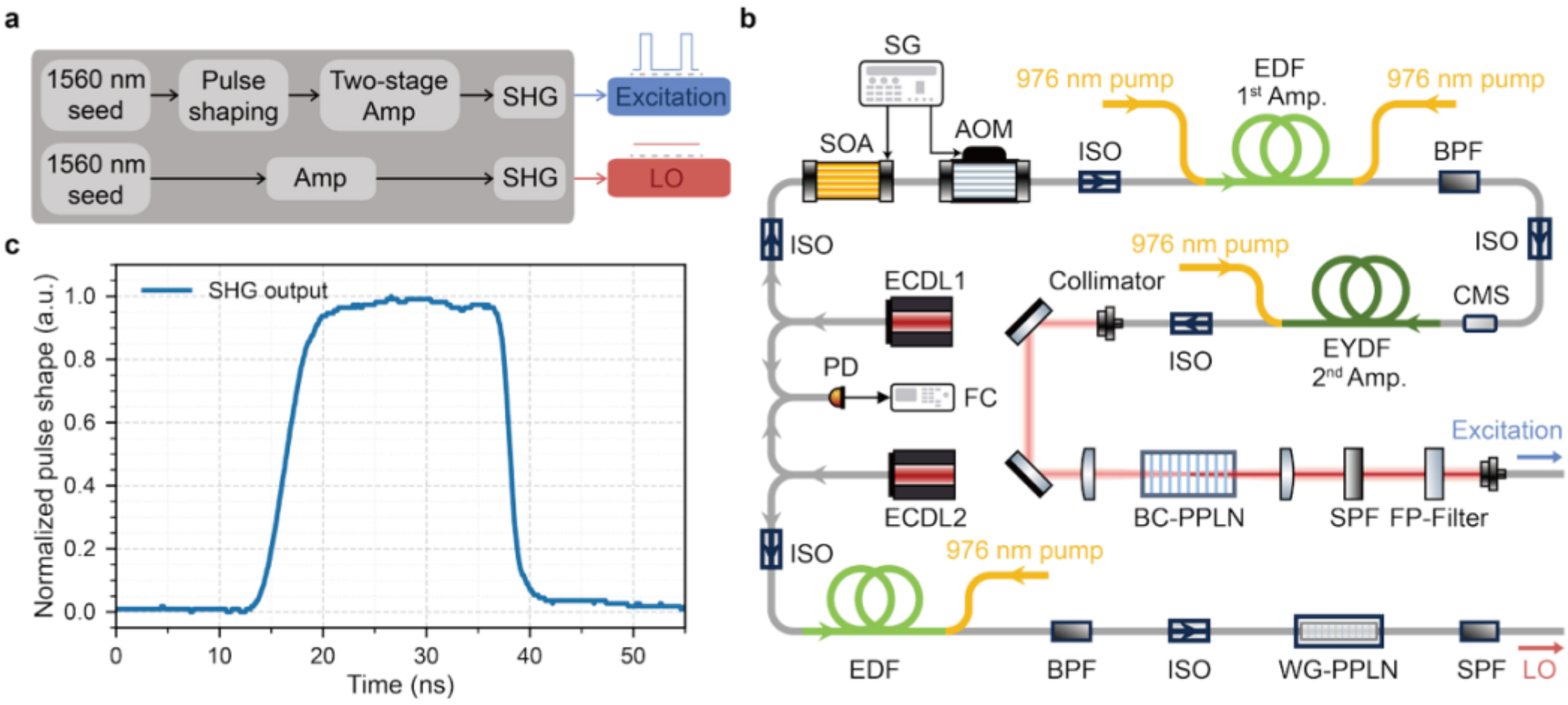

**Supplementary Fig. 4 Overview of the laser system**. **a**, Schematic of the laser system including a pulsed laser beam and a continuous-wave local oscillator (LO). **b**, Detailed experimental setup of the laser system. ECDL1 and ECDL2 are used as seed lasers for excitation and LO, respectively. ECDL1 output is chopped and temporally shaped using a cascaded semiconductor optical amplifier (SOA) and acousto-optic modulator (AOM) to reduce leading-edge intensity, compensating gain saturation in the subsequent MOPA chain for improved output pulse quality. A frequency counter (FC) is used to monitor the frequency difference between the two 1560 nm seed lasers with a photodetector (PD). ISO: fiber-optic isolator; EDF: Er-doped gain fiber; EYDF: Er/Yb co-doped double-cladding gain fiber; BPF: bandpass filter; CMS: cladding mode stripper; BC-PPLN: bulk periodically poled lithium niobate crystal; SPF: short-pass filter; FP-Filter: Fabry–Pérot etalon filter; WG-PPLN: waveguide periodically poled lithium niobate. **c**, Temporal profile of the generated 780-nm laser pulse with a duration of 25.6 ns.

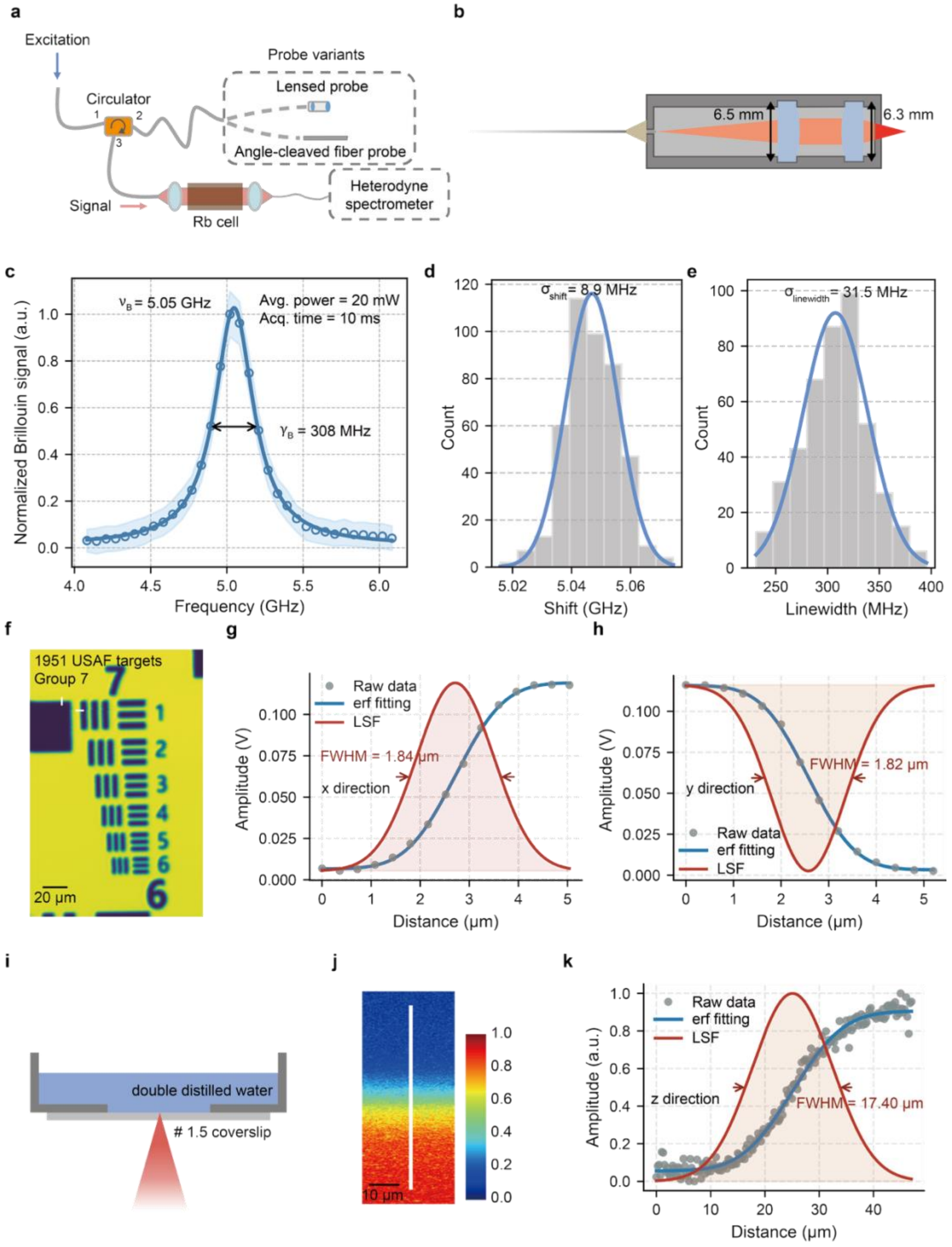

**Supplementary Fig. 5 Performance characterization of fiber-coupled PHBD configuration. a**, Schematic of the fiber-coupled PHBD setup. A pulsed excitation laser is routed through an optical circulator (port 1 to port 2) and delivered to the sample via the distal probe (lensed probe or angle-cleaved fiber probe). The backscattered Brillouin signal is collected by the same probe and routed to port 3 of the circulator. The output signal is then collimated and passed through a rubidium cell to suppress Rayleigh scattering and residual excitation laser leakage from port 1. The filtered signal is subsequently coupled into a heterodyne spectrometer for detection. **b**, Schematic of the lensed probe, which comprises a collimating lens and a focusing lens. **c**, Brillouin

spectra of water measured with an average power of 20 mW, exposure time of 10 ms, peak power of 18.9 W and pulse width of 16 ns. The open circles denote the mean of $n$ = 400 spectra, light blue shading indicates the standard deviation, and solid blue line is the corresponding Lorentzian fit. **d**,**e**, Histograms of the Lorentzian-fitted Brillouin shift (**d**) and linewidth (**e**) from the spectra in **c**, with blue curves representing Gaussian fits. **f**, Point-scanning image of Group 7 of a positive 1951 USAF resolution test target. **g**,**h**, Profiles of transmitted power detected by a photodetector along the x (**g**) and y (**h**) directions across the edge of the target pattern, as indicated by the white solid lines in **f**. Blue and red lines represent the error function fit and its corresponding gradient (line spread function), respectively. **i**, Schematic of the water–coverslip interface. **j**, Normalized Lorentzian-fitted Brillouin amplitude from double-distilled water along the axial direction. **k**, Brillouin signal amplitude profile along the z direction across the water–coverslip boundary, as indicated by the white solid line in **i**. Blue and red lines represent the error function fit and its corresponding gradient, respectively.

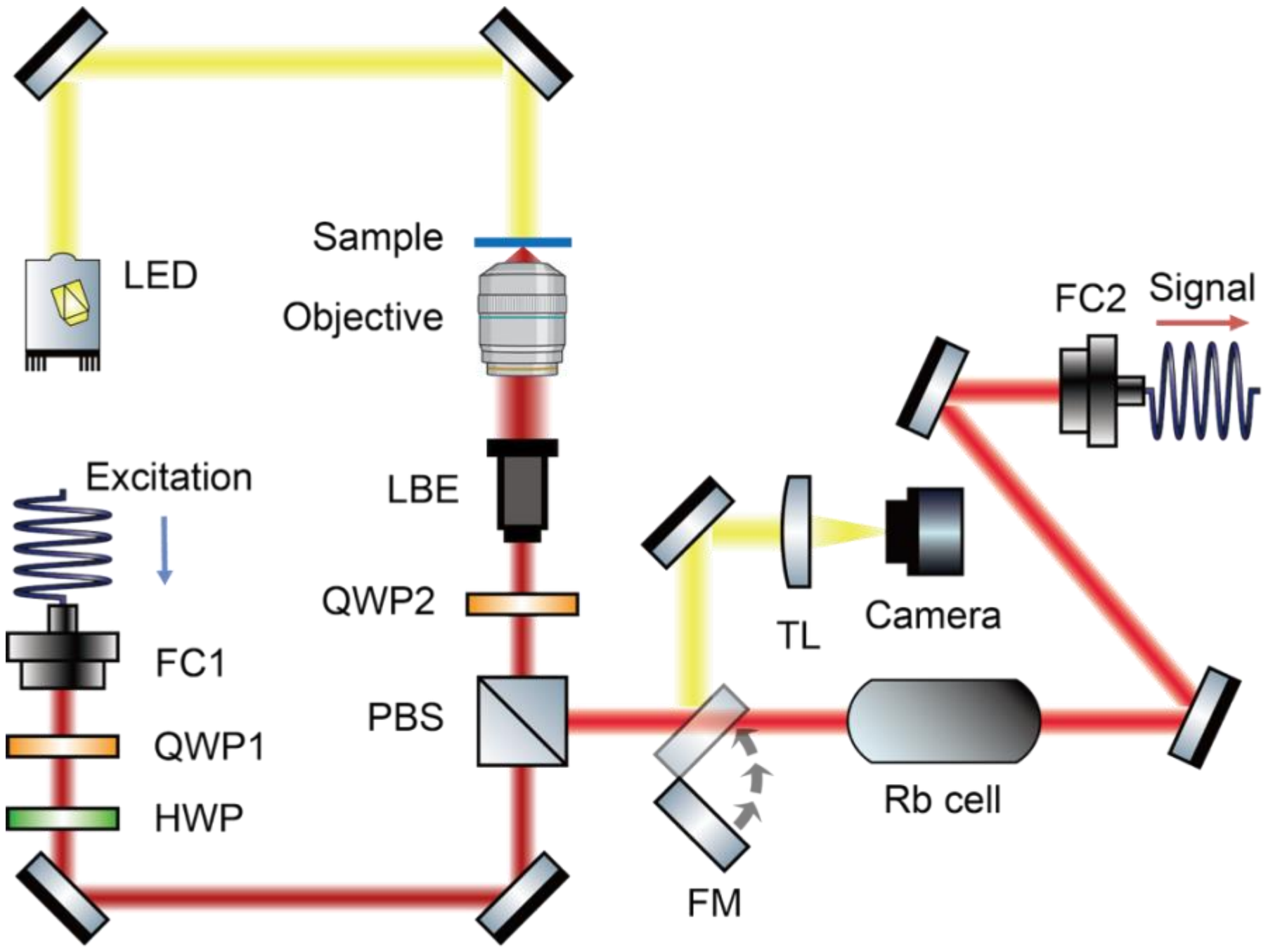

**Supplementary Fig. 6 Detailed setup of high-resolution microscopy configuration.** LED: light-emitting diode; FC: fiber collimator; QWP: quarter-wave plate; HWP: half-wave plate; LBE: laser beam expander; PBS: polarizing beam splitter; FM: flip mirror; TL: tube lens; Rb cell: rubidium cell. The quarter-wave plate and half-wave plate ensure the incident laser is transmitted through the PBS with minimal power loss. The optical paths for brightfield and Brillouin imaging modes are switched by the flip mirror. With the flip mirror up, the yellow beam carrying the bright-field information is imaged onto the camera. When the flip mirror is down, the Brillouin signal (red beam) passes through the Rb cell and is coupled into the signal fiber.

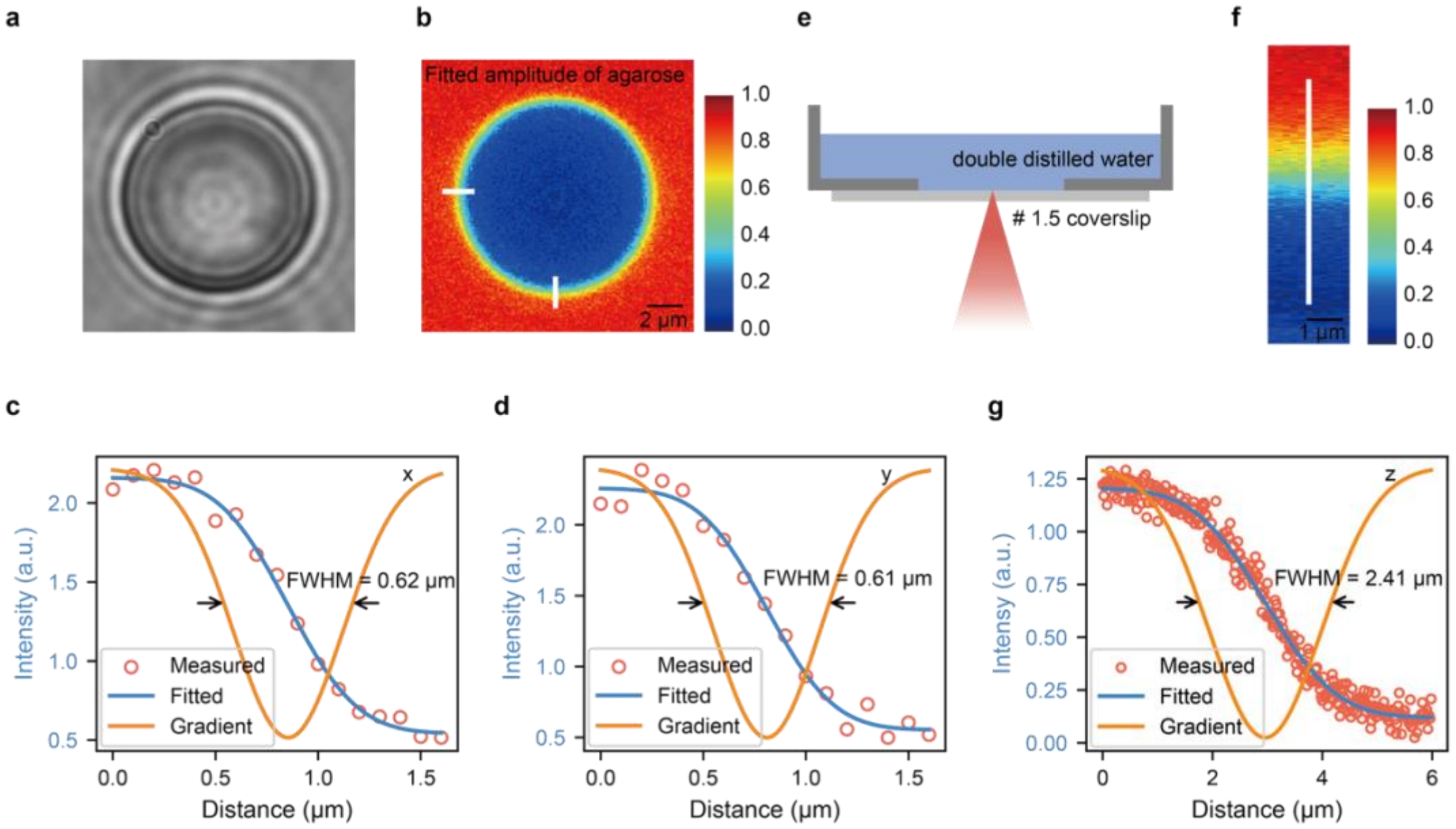

**Supplementary Fig. 7 Spatial resolution characterization of the free-space configuration with a 0.7 NA objective.** **a**, Brightfield image of a polymethyl methacrylate (PMMA) bead embedded in agarose. **b**, Normalized Lorentzian fitting amplitude of Brillouin signal from agarose. Since the Brillouin shift of PMMA lies outside the detected spectral range, a single Lorentzian fitting is applied to all measured spectra. **c**,**d**, Amplitudes of the Lorentzian fits along the x (**c**) and y (**d**) direction across the boundary of the PMMA bead, as indicated by the white solid lines in **b**. The blue and orange lines represent the error function fitting and their corresponding gradients, respectively. **e**, Schematic diagram of the water-glass interface. **f**, Normalized Lorentzian fitting amplitude of Brillouin signal from double-distilled water. The Brillouin shift of the coverslip also lies outside the detection range, ensuring that it does not affect the single Lorentzian fitting. **g**, Amplitude profile along the z direction across the boundary between water and the coverslip, as indicated by the white solid line in **f**. The blue and orange lines represent the error-function fit and its corresponding gradient, respectively.

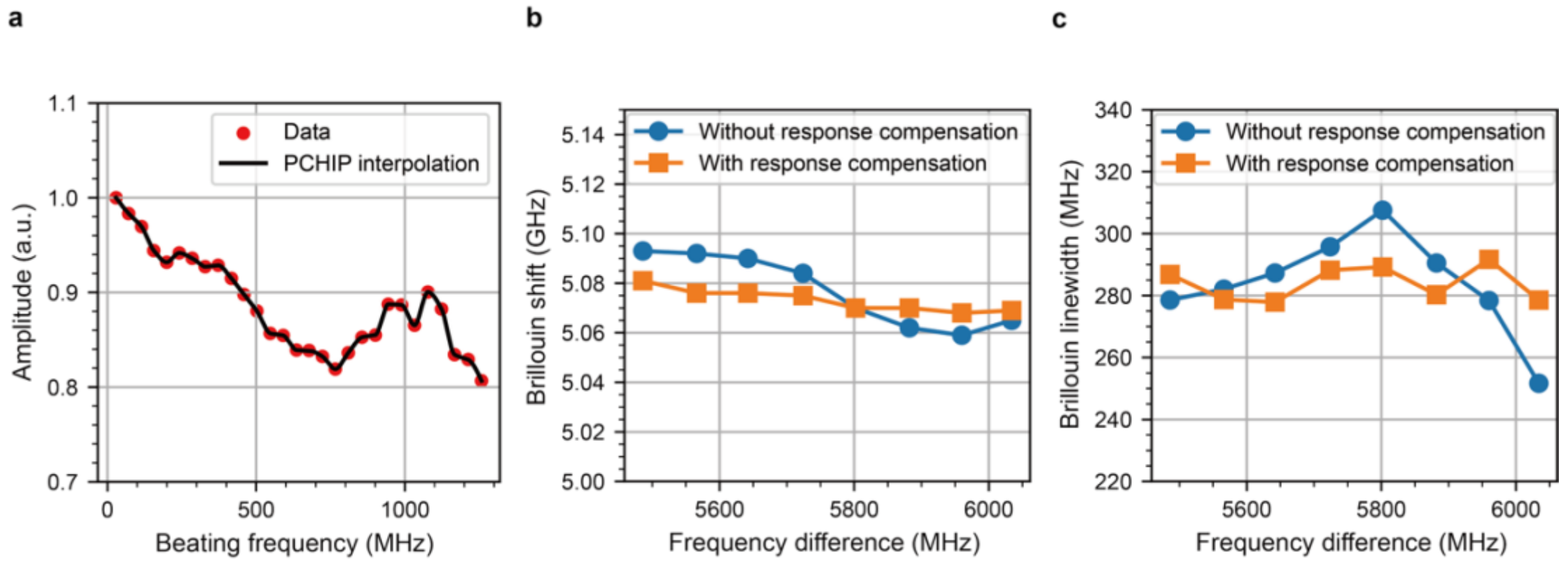

**Supplementary Fig. 8 Balanced photodetector frequency-response characterization and the spectrum compensation. a**, Balanced photodetector (BPD) response coefficient at different beat frequencies. The beat note is generated by two CW lasers, each with a power of 0.2 mW and detected using an electronic spectrum analyzer. The piecewise cubic Hermite interpolating polynomial (PCHIP) interpolation provides more accurate compensation when the FFT frequency points fall between measured frequency points. **b**, Comparison of the Brillouin shift of water for different excitation–LO frequency differences (before frequency doubling) with (squares) and without (circles) response compensation. **c**, Comparison of the Brillouin linewidth of water for different excitation–LO frequency differences (before frequency doubling) with (squares) and without (circles) response compensation. The water signals are measured under an average power of 35 mW and an exposure time of 20 ms. After compensating for the BPD response, the variations in both Brillouin shift and linewidth across different excitation–LO frequency differences are reduced, indicating improved spectral measurement accuracy. Because the BPD frequency response is stable, it was characterized once and then used for all subsequent measurements.

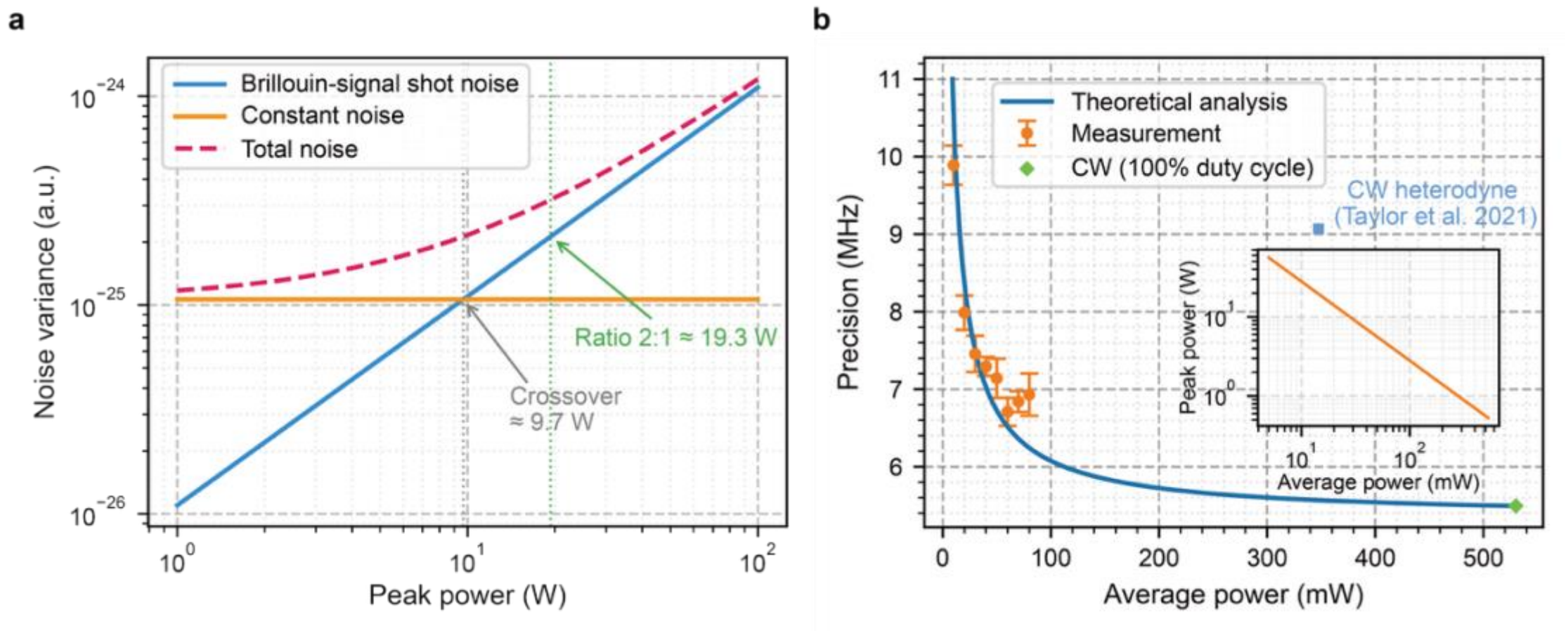

**Supplementary Fig. 9 Theoretical and experimental validation of pulse enhancement and analysis of noise contributions. a,** Contributions of Brillouin-signal shot noise and constant noise to the total noise variance as a function of peak excitation power. The two contributions are equal at a peak power of approximately 9.7 W. At approximately 19.3 W, the Brillouin-signal shot noise variance is twice the constant-noise variance, which we use as the criterion for entering the Brillouin-signal-shot-noise-limited regime. **b,** Brillouin shift precision measured in water as a function of average excitation power, with the corresponding peak power varied as shown in the inset. The solid blue curve represents the theoretical prediction, and the orange circles and error bars indicate the mean ± s.d. from five independent measurements, each comprising $n$ = 400 Brillouin spectra. The blue square denotes the previously reported CW heterodyne result from Supplementary Fig. 3b of Ref. [1]. Relative to this benchmark, pulsed excitation reduces the average power required to achieve comparable precision by 35-fold. The green diamond marks the theoretical CW limit of our system at a 100% duty cycle; relative to this limit, pulsed excitation reduces the required average power by 53-fold. All precision values were evaluated using the same acquisition time of 10 ms per spectrum.

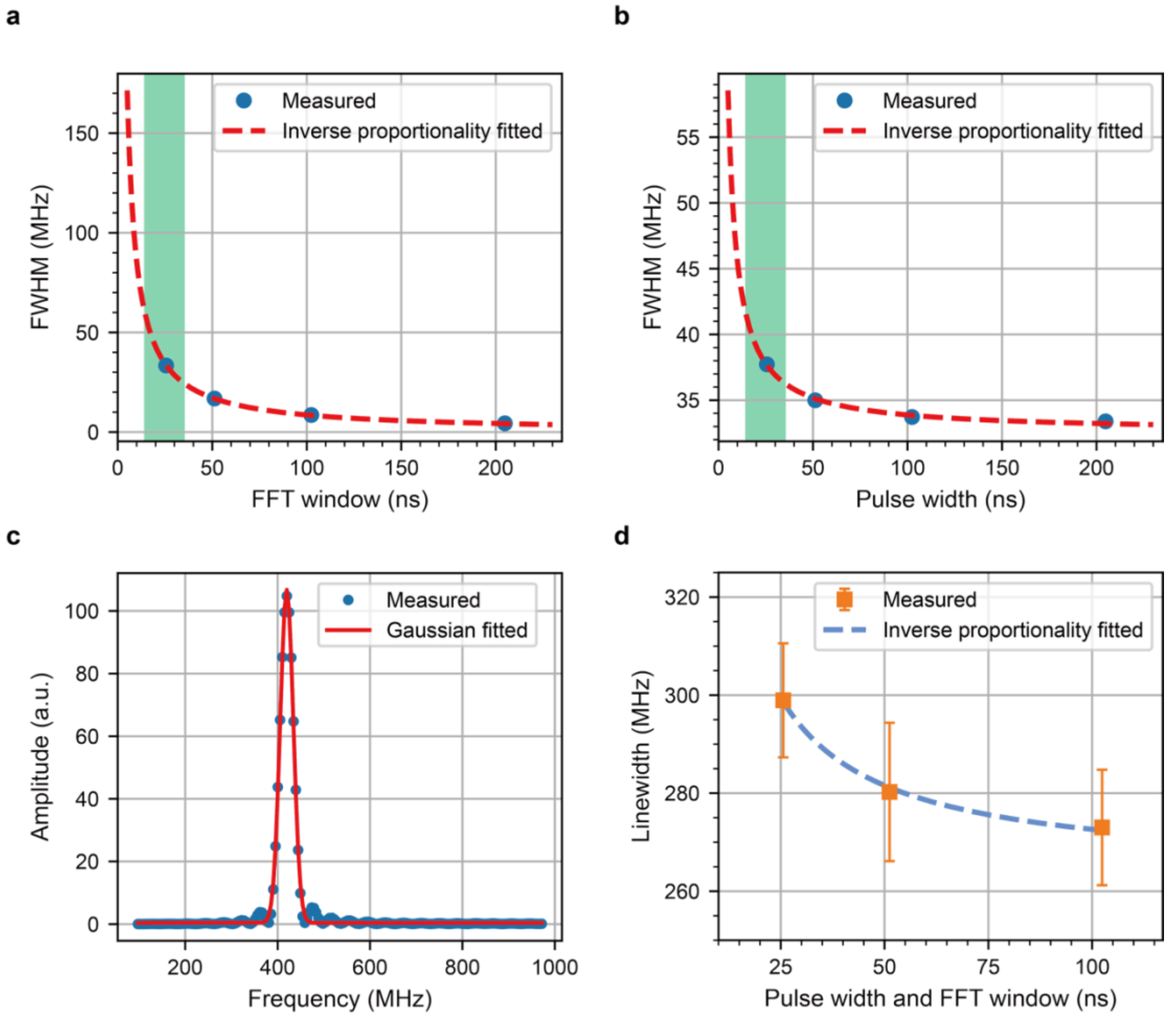

**Supplementary Fig. 10 Characterization of spectral resolution**. **a**, Measured full width at half maximum (FWHM) of the laser spectrum with a fixed pulse width of 204.8 ns as a function of fast Fourier transform (FFT) window duration. When the FFT window duration decreases below 15 ns, the measured FWHM of the laser increases sharply. **b**, Measured FWHM of the laser spectrum with a fixed FFT window duration of 25.6 ns as a function of pulse width. An offset arising from the 25.6-ns FFT window duration is subtracted from the data prior to inverse-proportional fitting. With a fixed FFT window, only a slight increase (~5 MHz) in FWHM is observed when the pulse width decreases from 204.8 ns to 25.6 ns. The main pulse widths used for PHBD are highlighted by the green shaded region. **c**, Typical laser spectrum measured with the heterodyne detection system. Both the pulse width and FFT window duration are set to 25.6 ns. Appropriate zero-padding is applied to improve the accuracy of Gaussian fitting. **d**, Measured Brillouin linewidths of double-distilled water with different pulse widths and corresponding FFT window durations with a low NA (0.12) objective. Error bars indicate the standard deviation of linewidths extracted from $n$ = 400 individual spectra. An offset of 263.7 MHz, corresponding to the intrinsic linewidth of double-distilled water, is subtracted from the measured linewidth prior to inverse-proportional fitting.

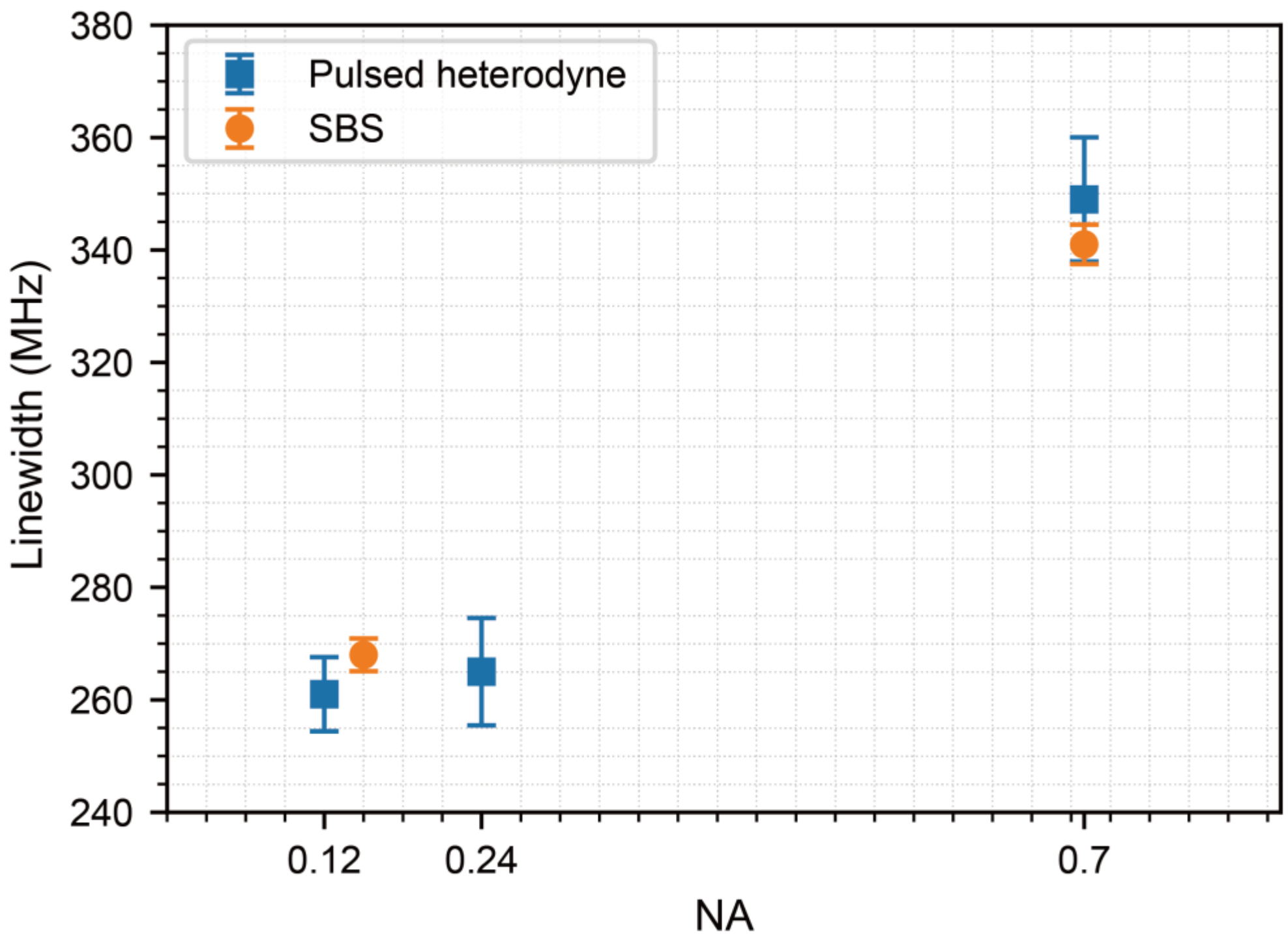

**Supplementary Fig. 11 Comparison of the Brillouin linewidth of water at different NAs in free-space PHBD configuration and stimulated Brillouin scattering (SBS) microscopy.** Brillouin linewidths of double-distilled water measured using free-space PHBD and quasi-pulsed SBS microscopy under different NAs. For the SBS measurements, the exposure times were set to 150 ms (for 0.15 NA) and 100 ms (for 0.7 NA), while the noise-equivalent power bandwidth of the lock-in amplifier was fixed at 800 Hz to minimize additional broadening from the detection system. The pulsed laser with a pulse width of 60 ns introduces approximately 2 MHz broadening. The average powers on sample for the probe and pump beams were set to 35 mW and 48 mW (for 0.15 NA) and 24 mW and 34 mW (for 0.7 NA), respectively, with an identical duty cycle of 5.2% to ensure a high SNR for accurate measurements. For the free-space PHBD measurements, the pulse width was set to 36 ns, introducing an effective FFT window that resulted in a theoretical overall spectral broadening of 27 MHz, including ~4 MHz from the finite pulse duration. The average power, peak power and exposure time were set to 45 mW, 22 W and 200 ms (for 0.12 NA); 10 mW, 21 W and 200 ms (for 0.24 NA); and 30 mW, 28 W and 200 ms (for 0.7 NA), respectively, achieving linewidth precisions of 6.6, 9.6 and 11.0 MHz. All spectral broadening contributions were subtracted to obtain the linewidths solely associated with NA-induced broadening shown in the figure. Error bars represent the standard deviation of linewidths extracted from $n$ = 400 individual spectra. The linewidths measured by the two methods agree well, confirming the accuracy of the spectra obtained with PHBD.

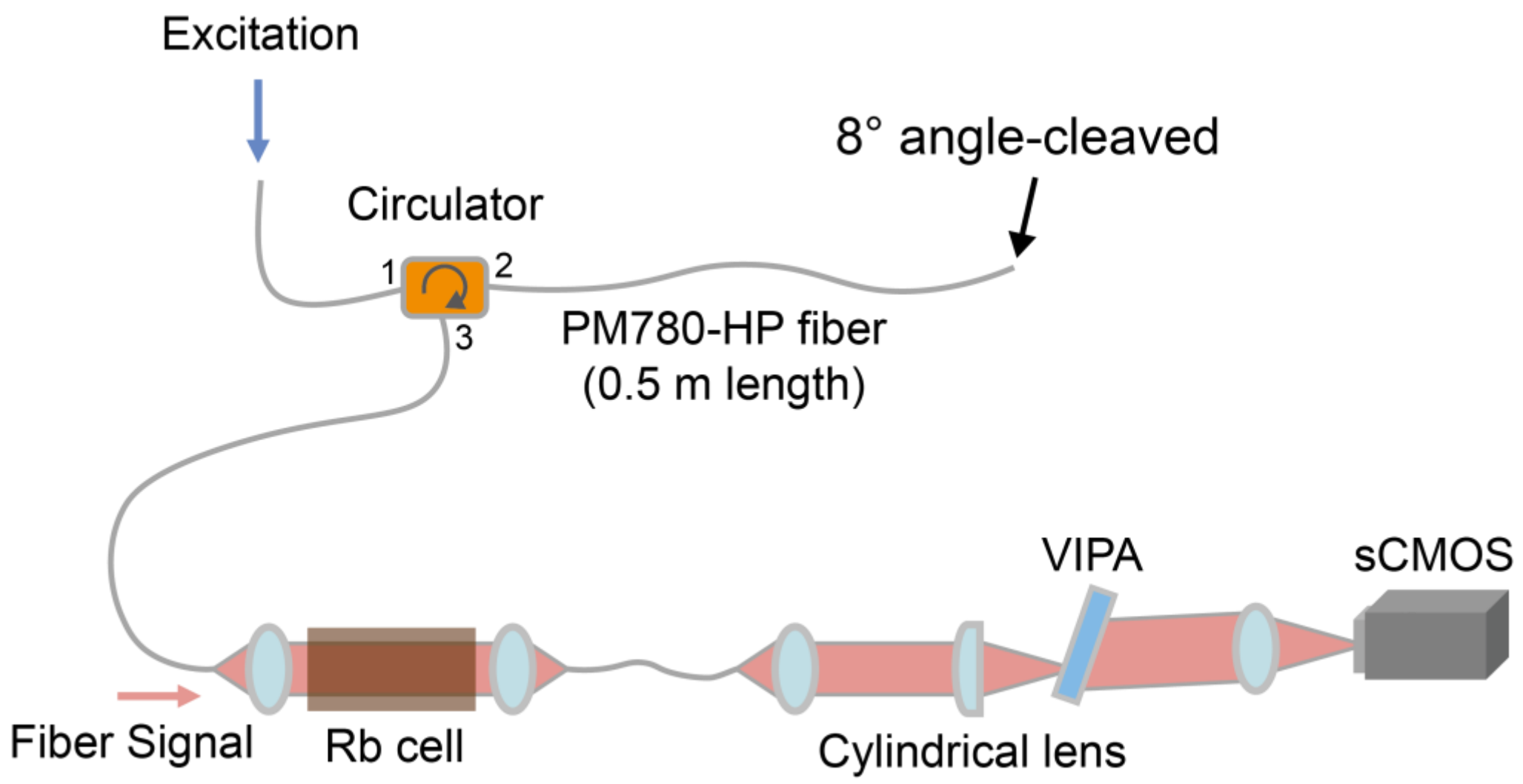

**Supplementary Fig. 12 Detailed optical setup for measuring fiber-generated background signal in a VIPA spectrometer.** The CW excitation laser is coupled into port 1 of an optical circulator and exits through port 2 into a 0.5-m PM780-HP optical fiber. Backscattered light from the fiber is directed via port 3 through a rubidium (Rb) absorption cell, which removes Rayleigh scattering and residual excitation laser leakage from port 1. The filtered signal is subsequently coupled into a polarization-maintaining fiber, spatially dispersed by a virtually imaged phased array (VIPA), and detected by an sCMOS camera. The fiber end at port 2 is angle-cleaved at 8° to minimize back-reflection.

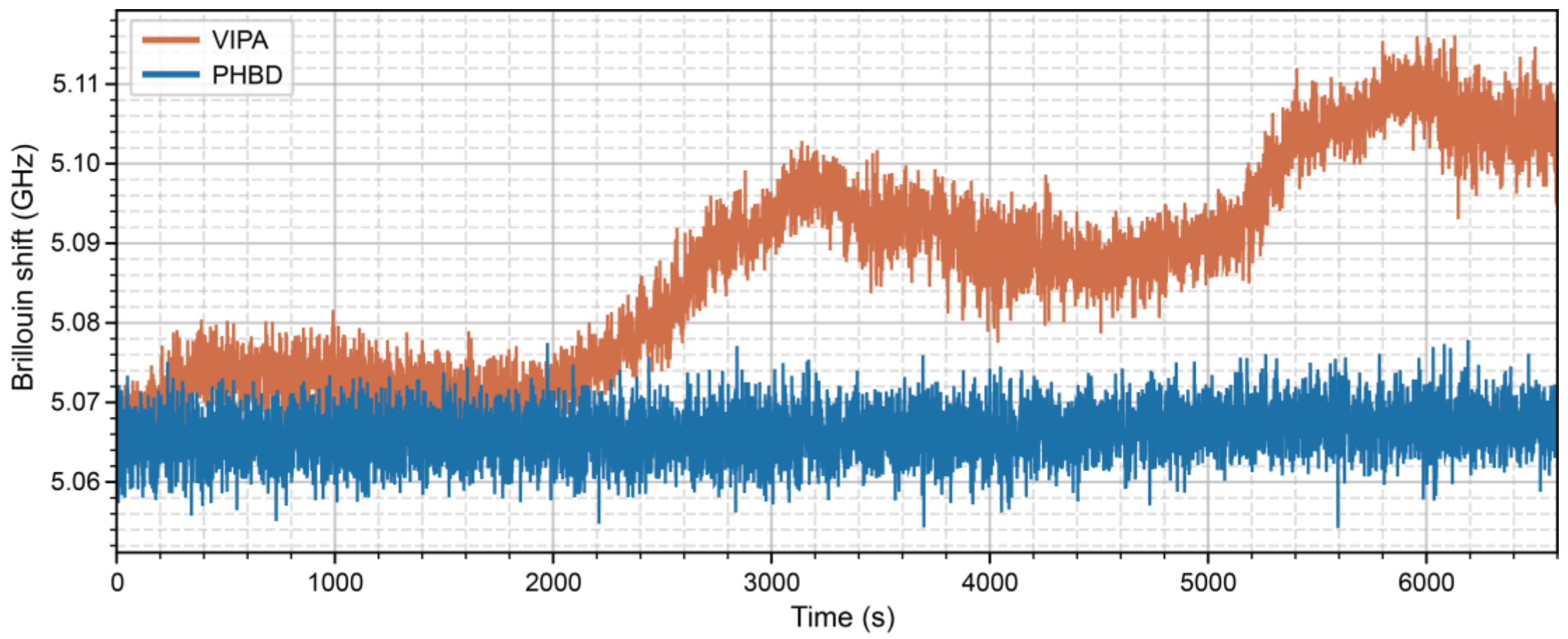

**Supplementary Fig. 13 Side-by-side comparison of measurement stability.** Long-term comparison of the measured Brillouin shift of water using PHBD and a VIPA spectrometer over 110 min. The experiments were performed in a temperature-controlled laboratory at 25 °C (variations of a few degrees) using the same low-NA objective (effective NA, 0.12); a single calibration was performed at the start. The ~2-MHz upward drift in the PHBD trace is attributed to excitation–LO frequency drift together with small temperature variations of the water sample.

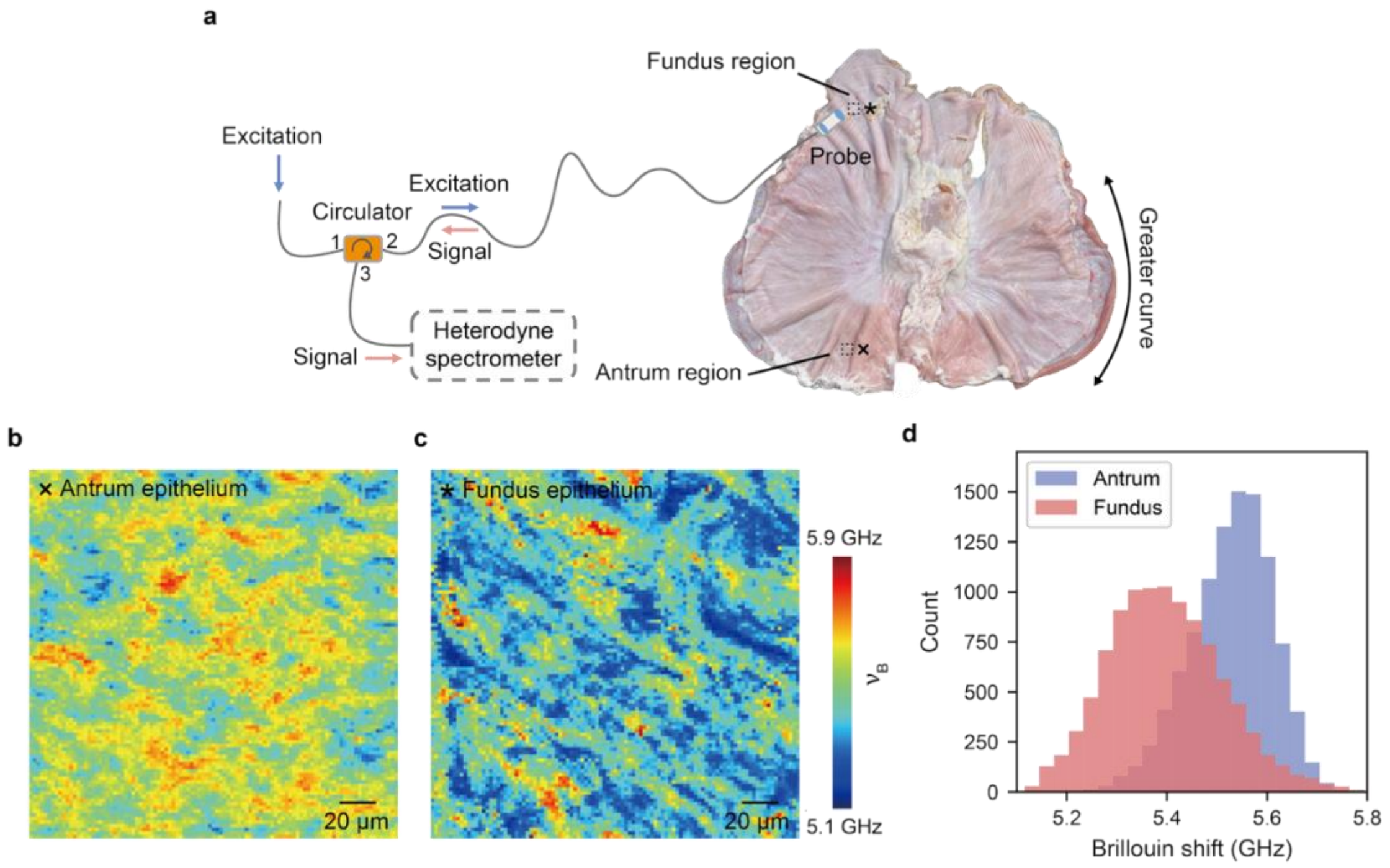

**Supplementary Fig. 14 Fiber-coupled Brillouin imaging of porcine stomach with PHBD. a,** Schematic of fiber-probe Brillouin imaging of a porcine stomach opened along the greater curvature. A lensed fiber probe was positioned at the gastric antrum and fundus to deliver excitation light and collect the backscattered Brillouin signal. **b**,**c**, Representative Brillouin-shift maps of the surface epithelium in the gastric antrum (**b**) and fundus (**c**), corresponding to the marked regions in **a**. The scan area is 202 µm × 202 µm for both regions, with a step size of 2 µm × 2 µm. Both maps are displayed using the same color scale. **d**, Pixel-level distributions of the Brillouin shifts in the antrum and fundus regions shown in **b** and **c**, revealing a higher Brillouin-shift distribution in the antrum of this representative stomach. The experiment was independently repeated using $n$ = 6 porcine stomachs, with a similar regional trend observed in all three specimens.

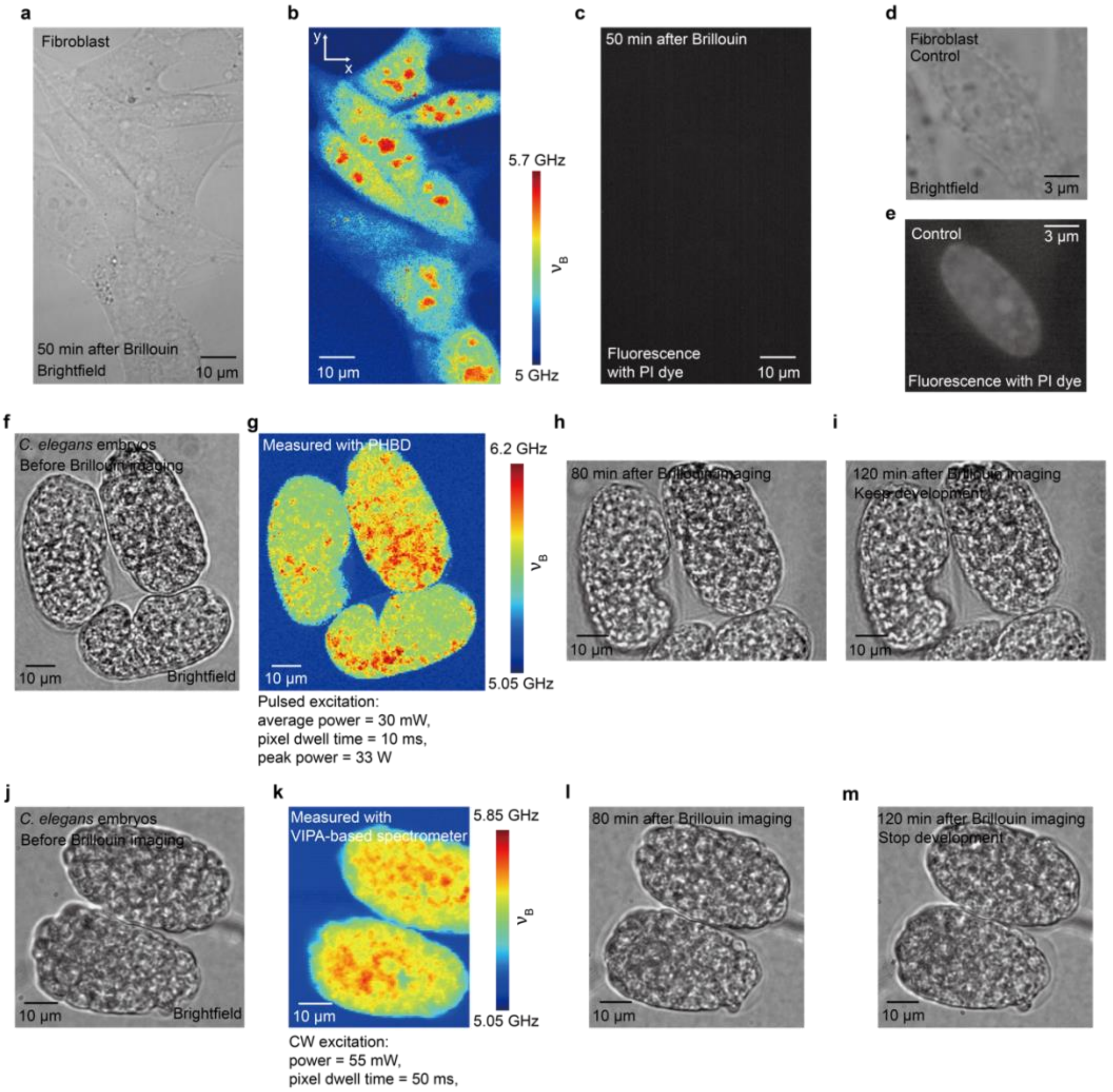

**Supplementary Fig. 15 Phototoxicity assessment. a,b,** Brightfield (**a**) and Brillouin shift (**b**) images of a fibroblast cell cluster; scan step, 0.25 μm × 0.25 μm. **c,** Fluorescence image of the same cluster 50 min after Brillouin imaging. The absence of PI fluorescence indicates the viability of the imaged cell cluster. **d,e,** Brightfield (**d**) and corresponding PI fluorescence (**e**) images of a positive-control fibroblast, confirming PI staining of membrane-compromised cells. **f–i**, Phototoxicity assessment using free-space PHBD. **f,** Brightfield image of three *C. elegans* embryos prior to Brillouin imaging. The embryo at the bottom is only partially visible due to the limited field of view of the brightfield modality. **g**, Brillouin shift image acquired with free-space PHBD (average power, 30 mW; pixel dwell time, 10 ms; peak power, 33 W). **h,i,** Brightfield images captured 80 min (**h**) and 120 min (**i**) after Brillouin imaging. Pronounced morphological changes observed across the time-lapse indicate continuous embryonic development ($n$ = 13). **j–m,** Phototoxicity assessment using continuous-wave (CW) excitation. **j,** Brightfield image of two *C. elegans* embryos prior to imaging. **k,** Brillouin shift image acquired using a VIPA spectrometer with a CW excitation laser (power, 55 mW; pixel dwell time, 50 ms). **l,m,** Brightfield images captured 80 min (**l**) and 120 min (**m**) after Brillouin imaging. In contrast, embryo morphologies showed no noticeable changes over the 40-minute interval between (**l**) and (**m**), indicating the arrest of development ($n$ = 9). These results demonstrate the capability of PHBD to image fragile embryos which is not available under CW illumination at 55 mW with a 50-ms pixel dwell time.

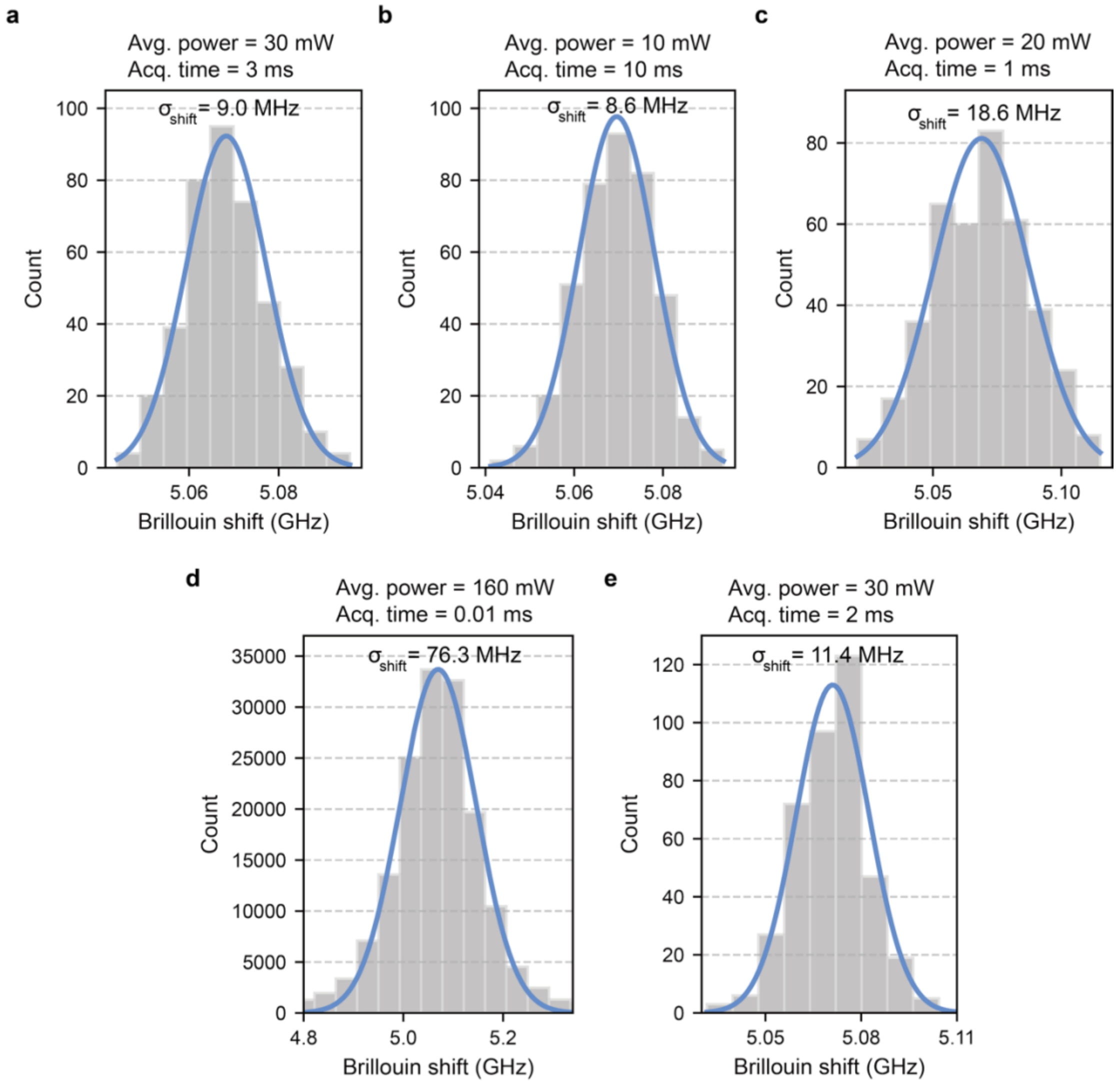

**Supplementary Fig. 16 Brillouin shift precision in water under different PHBD acquisition conditions. a–e**, Distributions of the measured Brillouin shifts in water. Gray bars show the histograms, blue curves show Gaussian fits and $\sigma_{shift}$ denotes the standard deviation of the spectral shift. For **a–c** and **e**, the peak power was fixed at 28 W, and the average power was adjusted by varying the pulse repetition rate to 30 kHz (**a**), 10 kHz (**b**), 20 kHz (**c**) and 30 kHz (**e**). For each condition, $n$ = 400 spectra were used to estimate $\sigma_{shift}$. For **d**, the peak power was 44 W and the repetition rate was 100 kHz; $n$ = 175,000 spectra were used to estimate $\sigma_{shift}$ due to the broader coverage of shifts. The pulse duration was fixed at 36 ns for all measurements. The average power and acquisition time for each condition are indicated above the corresponding panel.

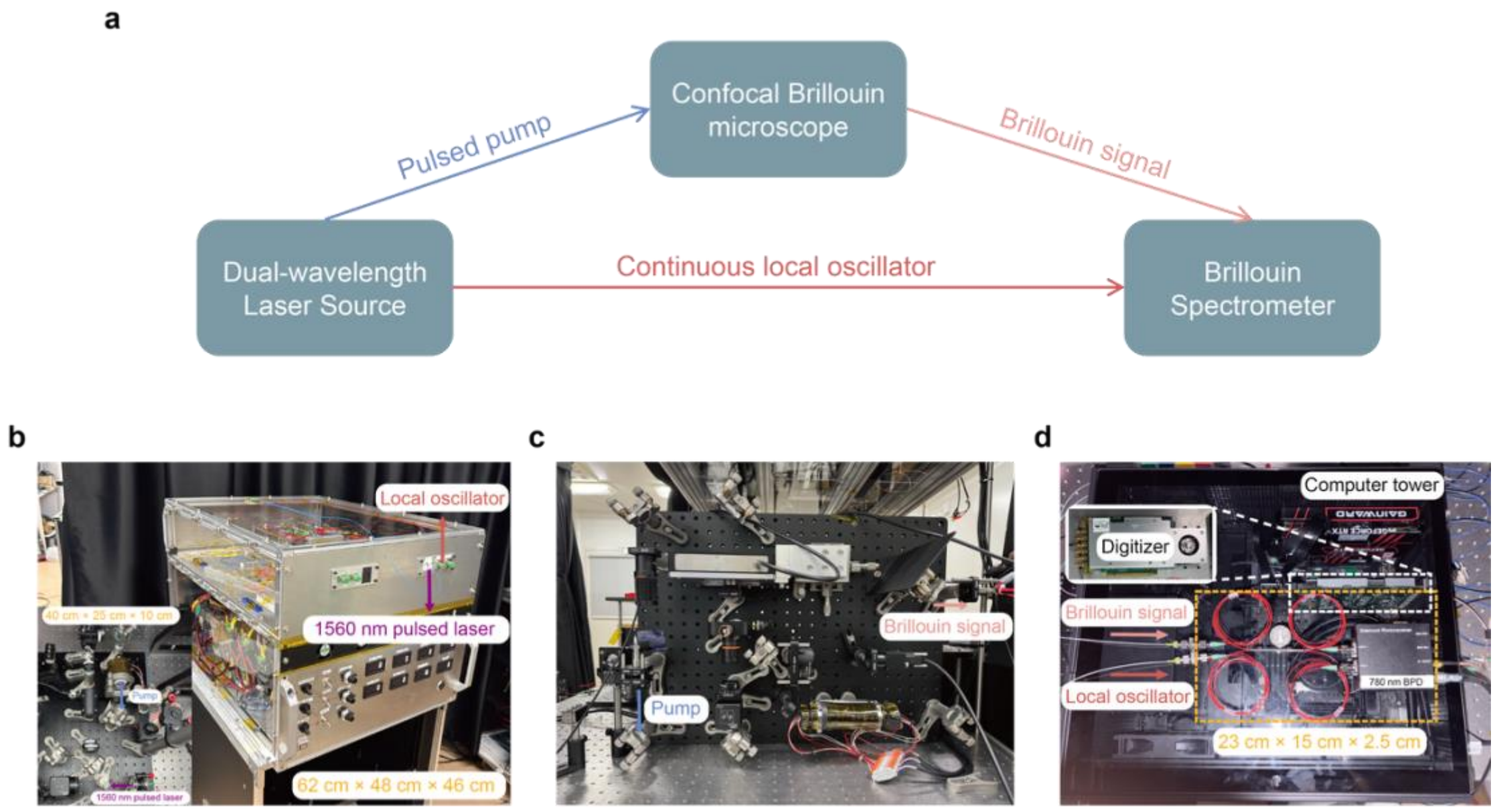

**Supplementary Fig. 17 Modular architecture of free-space PHBD configuration.** **a**, Schematic of the three main modules and their interconnections. **b–d**, Photographs of the dual-wavelength laser source (**b**), the microscope optical setup (**c**), and the heterodyne Brillouin spectrometer with the processing computer (**d**). The laser system (**b**) consists of a laser-source mainframe and a pulsed-laser second-harmonic generation (SHG) unit. The mainframe measures 62×48×46 $cm^3$ and contains three internal layers. The bottom layer houses seed-laser modules, single-mode-fiber coupled pump diodes, multimode-fiber coupled pump diode, their corresponding diode drivers, thermoelectric cooler (TEC) controllers, communication and control electronics, voltage converters, and air-cooling components. The second layer contains the fiber-amplifier chains for the short-wave infrared (SWIR) pulsed laser and LO seed. The top layer is dedicated to SHG of the local oscillator (LO) beam. The mainframe outputs the 1560 nm pulsed laser and the 780 nm CW LO beam through a PM1550-XP fiber pigtail and a PM780-HP fiber coupled FC/APC flange connector on the upper front panel, respectively. The upper back panel includes three PM1550-XP fiber-coupled FC/APC flange connectors for the seed laser output to the acousto-optic modulator (AOM) pulse shaper, the input of shaped seed pulses, and the seed-beating output. The lower front panel is equipped with switches and knobs for manual control, a display panel for real-time status monitoring, and a USB interface for communication and advanced control. The mainframe is powered by the 220 V AC mains via a power cord. The PM1550-XP fiber carrying the 1560 nm pulsed laser is connected to the SHG unit (shown in the inset) for frequency doubling, followed by spectral filtering and fiber coupling to the microscope. The main body of the microscope (**c**) is constructed on a breadboard and implements a compact confocal epi-illumination optical layout. It is worth noting that the optical configuration can be readily adapted to a commercial fluorescence microscope in either an upright or inverted geometry. The spectrometer (**d**) consists of only three components: a polarization-maintaining fiber coupler (FC), a balanced photodetector (BPD), and a digitizer. Because the digitizer is directly integrated into the processing computer tower, the effective volume of the heterodyne spectrometer is merely 23×15×2.5 $cm^3$, without requiring any additional shielding to block stray light.

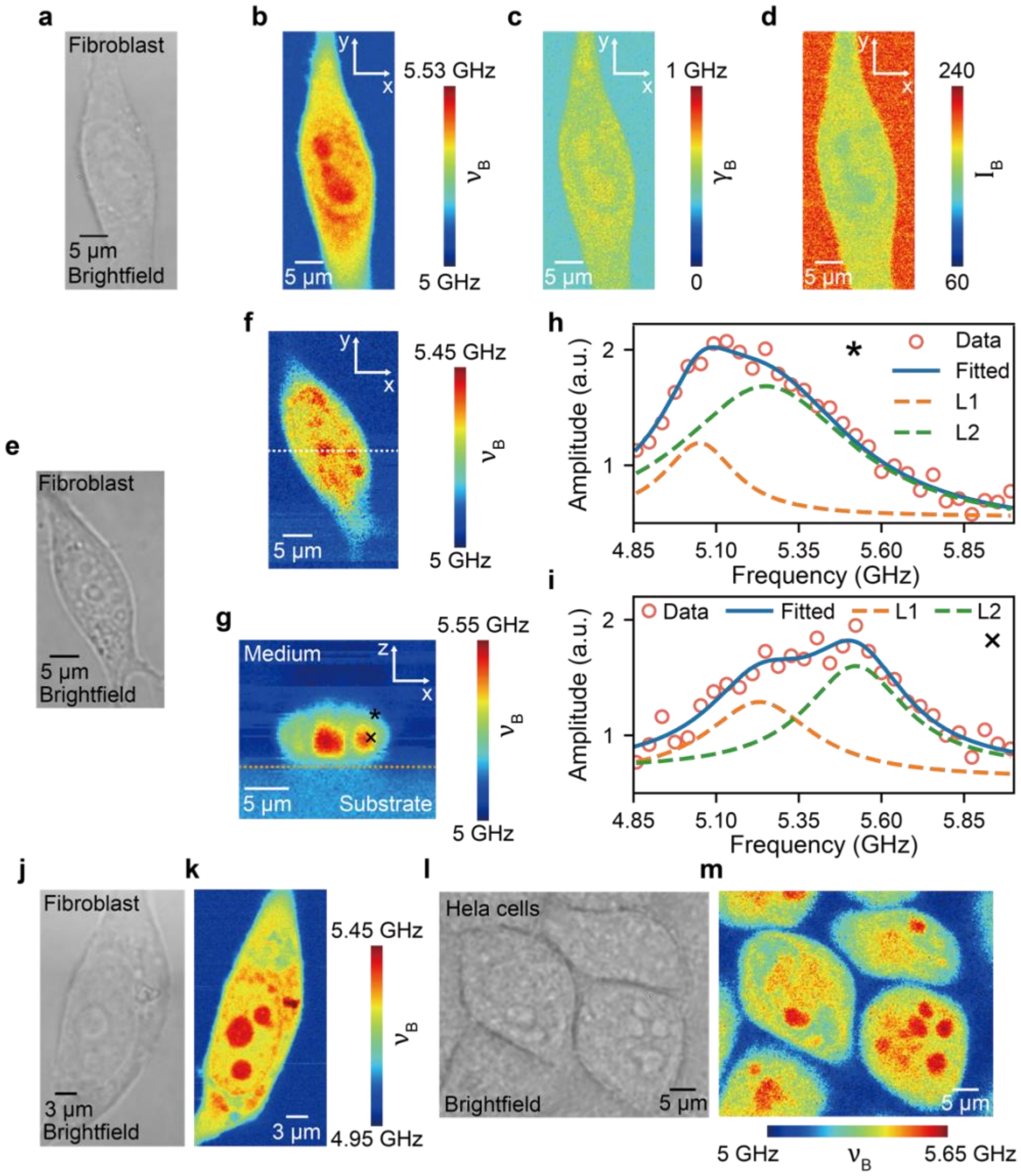

**Supplementary Fig. 18 High-spectral-resolution PHBD imaging of cells. a–d**, Brightfield image (**a**), Brillouin shift image (**b**), linewidth image (**c**) and intensity image (**d**) of a fibroblast. Each line in the intensity map was normalized by the corresponding signal intensity from the surrounding medium to compensate for power drift during imaging. **e,** Brightfield image of a fibroblast. **f,** Brillouin shift image of the fibroblast in the x-y plane. **g,** Brillouin shift image of the same cell in the x-z plane along the dashed line in **f**. The orange dashed line indicates the interface between the culture medium and the gel substrate. **h,i,** Representative Brillouin spectra acquired at the medium–cytoplasm boundary marked by the asterisk in **g** (**h**) and at the nucleoplasm–nucleolus boundary marked by the cross in **g** (**i**). PHBD resolves two closely spaced Brillouin components, with fitted frequency separations of 250 MHz in **h** and 291 MHz in **i**, respectively. Raw data are shown as red circles, double-Lorentzian fits as solid blue curves and the two fitted Lorentzian components, L1 and L2, as dashed curves. Offsets are added to L1 and L2 for visualization. **j,k,** Brightfield image (**j**) and Brillouin shift image (**k**) of a fibroblast acquired with an oil-immersion objective (UPLXAPO100XO, Olympus; effective NA, 1.22). **l,m,** Brightfield image (**l**) and Brillouin shift image (**m**) of a HeLa cell cluster. Except for **j** and **k**, all other images were acquired through a 0.7 NA objective (LUCPLFLN60X, Olympus).

## Supplementary Table 1. Positioning of PHBD among Brillouin modalities

| Method & reference | Detection & geometry | Spectral resolution | Shift precision | Avg. power | Acq. time / point | Epi (single-ended) | Fiber-compatible | Demonstrated samples |
|---|---|---|---|---|---|---|---|---|
| Confocal TFP<br>ref. 1,2 | Resonant interference<br>Epi / confocal | 100 MHz | – | 17 mW | 10 s | ✓ | ✗ | Cultured cells, biofilm |
| Confocal VIPA<br>ref. 3–5 | Spatial dispersive<br>Epi / confocal | ~300 MHz | <10 MHz | 3–7 mW | 50–100 ms | ✓ | ✗ | Cultured cells, cornea, plant tissue |
| Multiplexed VIPA line-scan / full-field<br>ref. 6–8 | Spatial Dispersive<br>Orthogonal / transmission | 0.25–1 GHz | 10–83 MHz | 18–370 mW | 0.02–1 ms | ✗ | ✗ | Embryos, spheroid, zebrafish |
| SBS<br>CW / pulsed<br>ref. 9–11 | Nonlinear pump–probe<br>Dual-objective / transmission | ~100 MHz | <10 MHz | 27–265 mW | 0.2–20 ms | ✗ | ✗ | C. elegans, cultured cells, zebrafish, organoids, embryos |
| CW heterodyne<br>ref. 12 | Coherent beating<br>Epi / confocal | 3.1 MHz | ~9 MHz | 345 mW | 10 ms | ✓ | ✗ | Cultured cells, fibrin fiber network |
| Fiber-optic TFP<br>ref. 13,14 | Resonant interference<br>Hollow-core fiber or dual-fiber / confocal | 276 MHz | 20 MHz | 10 mW | 20–30 s | ✓ | ✓ | Phantoms |
| Fiber-optic time-resolved analysis<br>ref. 15,16 | Transient interference<br>Epi / opto-acoustic transducer | 160 MHz | 7 MHz* | 15 mW | 2.5 s+ | ✓ | ✓ | Cultured cells, C. elegans |
| **PHBD**<br>**this work** | **Pulsed coherent beating**<br>**Epi / PM-fiber / confocal** | **25 MHz** | **8.9–9.0 MHz** | **20–30 mW** | **3–10 ms** | ✓ | ✓ | **Cultured cells, plant tissue, animal tissues, *in vivo* mouse deep brain** |

TFP: tandem Fabry-Pérot; VIPA: virtually imaged phased array; CW: continuous wave

* The reported 7 MHz shift precision in Ref. [16] was obtained from the frequency measurement of the entire time-domain ToF signal, rather than from depth-resolved voxel-wise frequency measurements. Therefore, it should be associated with the full x-y acquisition time, not the effective per-voxel time obtained by dividing the full ToF acquisition time by the number of depth-resolved z voxels.

+ This value denotes the total time-of-flight (ToF) acquisition time before voxel-wise normalization, used to achieve a shift precision of 7 MHz.

**Supplementary Table 2. Summary of PHBD measurement configurations for biological samples**

| | Average power (mW) | Acquisition time (ms) | Peak power (W) | Pulse width (ns) | Detection bandwidth (GHz) | Configuration |
|---|---|---|---|---|---|---|
| Fig. 3d,e (brain) | 25 | 50 | 17.9 | 14 | 2.5 | *Fiber – coupled (bare fiber probe)* |
| Fig. 3g (stomach) | 40 | 20 | 20.8 | 16 | 2.5 | *Fiber – coupled (bare fiber probe)* |
| Fig. 4b (plant root)<br>Fig. 4g,h (kidney) | 30 | 10 | 46.9 | 16 | 2.5 | *Free – space* |
| Fig. 4d (plant root) | 30 | 100 | 46.9 | 25.6 | 2.5 | *Free – space* |
| SI Fig. 14b,c (stomach) | 40 | 20 | 20.8 | 16 | 2.5 | *Fiber – coupled (lensed probe)* |
| SI Fig. 15b (cell)<br>SI Fig. 18f,m (cell) | 30 | 10 | 32.6 | 25.6 | 1.25 | *Free – space* |
| SI Fig. 15g (embryo) | 30 | 10 | 32.6 | 25.6 | 2.5 | *Free – space* |
| SI Fig. 18a–d (cell) | 30 | 10 | 33.5 | 25.6 | 2.5 | *Free – space* |
| SI Fig. 18g (cell) | 30 | 60 | 32.6 | 25.6 | 1.25 | *Free – space* |
| SI Fig. 18k (cell) | 38 | 10 | 42.4 | 25.6 | 2.5 | *Free – space* |

## Supplementary Note 1. Principle of pulse enhancement

To establish the theoretical basis of pulse enhancement, we first derive the signal-to-noise ratio (SNR) for our heterodyne Brillouin microscopy configuration. A general noise model for heterodyne detection was developed in Ref. [2]. Here, we adapt this framework to our experimental configuration and derive the signal, associated noise contributions and resulting SNR.

The spontaneous Brillouin field and the local-oscillator (LO) field are represented by the following complex electric fields:

$$\tilde{\boldsymbol{E}}_{Sp}(t) = \hat{x}E_{Sp}(t)e^{i[2\pi f_{Sp}t+\varphi_{Sp}(t)]} \tag{1}$$

$$\tilde{\boldsymbol{E}}_{LO}(t) = \hat{x}E_{LO}e^{i2\pi f_{LO}t} \tag{2}$$

where the tilde denotes the complex representation of the electric field (corresponding physical fields are given by $\tilde{\boldsymbol{E}}_{Sp/LO}^{phys}(t) = Re[\tilde{\boldsymbol{E}}_{Sp/LO}(t)]$), $E_{Sp}$ and $E_{LO}$ are the field amplitudes, $f_{Sp}$ and $f_{LO}$ are the carrier frequencies of the Brillouin signal and LO, respectively, and $\varphi_{Sp}$ is the phase difference between them. In our polarization-maintaining configuration, both optical fields are aligned along the same polarization direction $\hat{x}$; therefore, the vector character can be omitted in the following derivation.

After interference in a 2 × 2 fiber coupler with 50:50 splitting ratio, the complex fields at the two output ports are given by

$$\begin{bmatrix}\tilde{E}_{C1}(t)\\ \tilde{E}_{C2}(t)\end{bmatrix} = \begin{bmatrix}\frac{1}{\sqrt{2}} & i\frac{1}{\sqrt{2}}\\ i\frac{1}{\sqrt{2}} & \frac{1}{\sqrt{2}}\end{bmatrix}\begin{bmatrix}\tilde{E}_{Sp}(t)\\ \tilde{E}_{LO}(t)\end{bmatrix} \tag{3}$$

Following square-law photodetection and averaging over the optical oscillations, the optical powers incident on the two photodiodes are

$$\begin{aligned} P_{C1}(t) &= A_{eff}\epsilon_0 c\langle\{Re[\tilde{E}_{C1}(t)]\}^2\rangle_{opt} = \frac{A_{eff}\epsilon_0 c}{2}\left|\tilde{E}_{C1}(t)\right|^2 \\ &= \frac{A_{eff}\epsilon_0 c}{4}E_{Sp}^2(t) + \frac{A_{eff}\epsilon_0 c}{4}E_{LO}^2 + \frac{A_{eff}\epsilon_0 c}{2}E_{Sp}(t)E_{LO}sin\left[2\pi f_d t + \varphi_{Sp}(t)\right] \end{aligned} \tag{4}$$

$$\begin{aligned} P_{C2}(t) &= A_{eff}\epsilon_0 c\langle\{Re[\tilde{E}_{C2}(t)]\}^2\rangle_{opt} = \frac{A_{eff}\epsilon_0 c}{2}\left|\tilde{E}_{C2}(t)\right|^2 \\ &= \frac{A_{eff}\epsilon_0 c}{4}E_{Sp}^2(t) + \frac{A_{eff}\epsilon_0 c}{4}E_{LO}^2 - \frac{A_{eff}\epsilon_0 c}{2}E_{Sp}(t)E_{LO}sin\left[2\pi f_d t + \varphi_{Sp}(t)\right] \end{aligned} \tag{5}$$

where $A_{eff}$ is the effective optical mode area on PD, $\epsilon_0$ is the vacuum permittivity, $c$ is the vacuum light speed, $f_d = f_{Sp} - f_{LO}$ is the frequency difference between Brillouin signal and LO, i.e., the beat frequency. The differential output of BPD is then expressed as

$$s(t) = \mathcal{R}_p A_{eff} \epsilon_0 c E_{Sp}(t) E_{LO} sin[2\pi f_d t + \varphi_{Sp}(t)]$$
$$= 2\mathcal{R}_p \sqrt{P_{Sp}(t) P_{LO}} sin[2\pi f_d t + \varphi_{Sp}(t)] \quad (6)$$

where $\mathcal{R}_p$ denotes the responsivity of the detector, and $P_{Sp}$ and $P_{LO}$ are the optical powers of the spontaneous Brillouin signal and LO, respectively.

In heterodyne detection, the measured signal is accompanied by an approximately white noise component $e(t)$, arising predominantly from thermal noise and shot noise. The power spectral density of the total detection noise can be expressed as

$$\sigma_e^2 \approx \sigma_T^2 + 2q\mathcal{R}_p P_{LO} \quad (7)$$

where $\sigma_T^2$ is the thermal noise contribution and $q$ is the elementary charge. Because the LO power is typically in the milliwatt range, whereas the spontaneous Brillouin signal power is at the femtowatt level, shot noise associated with the Brillouin signal is negligible compared with that associated with the LO.

After digitization at a sampling rate of $f_s$, the sampled BPD output, including both signal and noise, is

$$r(n) = s(n) + e(n), \qquad n = 1,2,3, \ldots, N_s, \quad (8)$$

where $N_s$ is the total number of sampled points. To obtain the spectrum, a discrete fast Fourier transform (FFT) is applied to $r(n)$, yielding

$$\mathrm{R(k) = S(k) + E(k)}, \qquad \mathrm{k = 1,2,3, \ldots, N_s}, \quad (9)$$

Because spontaneous Brillouin measurements are analyzed in the spectral-power domain, we consider the normalized squared magnitude of the frequency-domain response. Taking the squared magnitude of both sides of equation (9) yields

$$\frac{|R(k)|^2}{N_s} = \frac{|S(k)|^2}{N_s} + \frac{|E(k)|^2}{N_s} + \frac{2S_{Re}(k)E_{Re}(k) + 2S_{Im}(k)E_{Im}(k)}{N_s} \quad (10)$$

To derive the overall SNR, the three terms on the right-hand side of equation (10) are considered separately.

The mean detected signal is determined by the first term on the right-hand side of equation (10). Applying Parseval's theorem directly, without introducing an auxiliary function as in Ref. [2], gives the correspondence between the time-domain signal power and its frequency-domain representation:

$$\sum_{k=1}^{N_s} \overline{\frac{|S(k)|^2}{N_s}} = \sum_{n=1}^{N_s} \overline{|s(n)|^2} = 2N_s \mathcal{R}_p^2 P_{LO} \bar{P}_{Sp}(t) \quad (11)$$

For a single-sided energy spectrum, assuming that the measured Brillouin spectrum follows a Lorentzian line shape and spectral sampling is dense, the signal energy in equation (11) can be written as

$$\sum_{k=1}^{\frac{N_s}{2}} \frac{\overline{|S(k)|^2}}{N_s} = \sum_{k=1}^{\frac{N_s}{2}} \frac{\overline{|S(k_d)|^2}}{N_s} \frac{\gamma_B^2}{\gamma_B^2 + 4[(k-k_d)\delta]^2} \tag{12}$$

where $\gamma_B$ denotes the linewidth of the Lorentzian line shape, $k_d$ is the spectral-bin index corresponding to the heterodyne beat frequency $f_d$ in the one-sided spectrum, and $\delta = \frac{f_s}{N_s}$ is the frequency spacing between adjacent spectral bins. Using one-based indexing, $k_d$ is given by

$$k_d = \frac{f_d}{f_s/N_s} + 1 \tag{13}$$

The sampling bandwidth is selected to be substantially broader than the Brillouin linewidth so that the full spectrum is captured. For sufficiently large $N_s$, the discrete summation in equation (12) approaches the corresponding Lorentzian integral:

$$\lim_{\delta \to 0} \delta \sum_{k=1}^{\frac{N_s}{2}} \frac{\overline{|S(k_d)|^2}}{N_s} \frac{\gamma_B^2}{\gamma_B^2 + 4(k-k_d)^2} = \frac{\pi}{2} \frac{|S(k_d)|^2}{N_s} \gamma_B \tag{14}$$

Combining equations (11), (12) and (14) gives the signal spectral density at the Lorentzian peak:

$$\frac{|S(k_d)|^2}{N_s} = \frac{2f_s}{\pi\gamma_B} \mathcal{R}_p^2 P_{LO} P_{Sp}(t) \tag{15}$$

Its expectation value is therefore

$$\frac{\overline{|S(k_d)|^2}}{N_s} = \frac{2f_s}{\pi\gamma_B} \mathcal{R}_p^2 P_{LO} \bar{P}_{Sp} \tag{16}$$

where $\bar{P}_{Sp}$ denotes the mean spontaneous Brillouin power.

Because the Brillouin signal is extremely weak in microscopy applications, fluctuation in the signal-only spectral term, $\frac{|S(k)|^2}{N_s}$ is negligible compared with the other noise contributions. The dominant contributions therefore arise from the detector-noise term and the signal–noise cross term, corresponding to the second and third terms on the right-hand side of equation (10), respectively. Since $E(k)$ is the discrete Fourier spectrum of Gaussian white noise, the expected normalized noise spectral power at $k_d$ is

$$\frac{\overline{|E(k_d)|^2}}{N_s} = \sigma_e^2 \frac{f_s}{2} \tag{17}$$

In PHBD, the BPD bandwidth is at least as large as the Nyquist bandwidth of the digitizer, $\frac{f_s}{2}$. The effective noise bandwidth is therefore set by the digitizer and is already included in equation (17). For Gaussian white noise, the variance of the normalized spectral-noise power is given by Ref. [3]:

$$D\left\{\frac{|E(k_d)|^2}{N_s}\right\} \approx \left(\frac{\overline{|E(k_d)|^2}}{N_s}\right)^2 = \frac{1}{4}\sigma_e^4 f_s^2 \tag{18}$$

A second noise contribution arises from the signal–noise cross term. At $k = k_d$, its variance can be evaluated as

$$\begin{aligned} &D\left\{\frac{2S_{Re}(k_d)E_{Re}(k_d) + 2S_{Im}(k_d)E_{Im}(k_d)}{N_s}\right\} \\ &= \frac{4}{N_s^2}\overline{[S_{Re}(k_d)E_{Re}(k_d) + S_{Im}(k_d)E_{Im}(k_d)]^2} \\ &= \frac{4}{N_s^2}\left[\overline{S_{Re}^2(k_d)}\ \overline{E_{Re}^2(k_d)} + \overline{S_{Im}^2(k_d)}\ \overline{E_{Im}^2(k_d)}\right] \\ &= 2\frac{\overline{|S(k_d)|^2}}{N_s}\frac{\overline{|E(k_d)|^2}}{N_s} = \frac{2f_s^{\ 2}}{\pi\gamma_B}\mathcal{R}_p^2 P_{LO}\bar{P}_{Sp}\sigma_e^2 \end{aligned} \tag{19}$$

In deriving equation (19), we used

$$\overline{2S_{Re}\ (k_d)E_{Re}\ (k_d)} = \overline{2S_{Im}\ (k_d)E_{Im}\ (k_d)} = 0 \tag{20}$$

$$\overline{2S_{Re}(k_d)E_{Re}(k_d)S_{Im}(k_d)E_{Im}(k_d)} = 0 \tag{21}$$

$$\overline{S_{Re}^2(k_d)} = \overline{S_{Im}^2(k_d)} \tag{22}$$

$$\overline{E_{Re}^2(k_d)} = \overline{E_{Im}^2(k_d)} \tag{23}$$

These relations follow from the random phases and zero-mean, statistically independent quadratures of the Fourier-domain spontaneous Brillouin signal and Gaussian white noise.

Combining the signal and noise contributions derived above, the SNR at the Brillouin spectral peak is

$$SNR_p = \frac{\frac{\overline{|S(k_d)|^2}}{N_s}}{\sqrt{D\left\{\frac{2S_{Re}(k_d)E_{Re}(k_d) + 2S_{Im}(k_d)E_{Im}(k_d)}{N_s}\right\} + D\left\{\frac{|E(k_d)|^2}{N_s}\right\}}}$$
$$= \frac{\frac{2f_s}{\pi\gamma_B}\mathcal{R}_p^2 P_{LO}\bar{P}_{Sp}}{\sqrt{\frac{2f_s^2}{\pi\gamma_B}\mathcal{R}_p^2 P_{LO}\bar{P}_{Sp}\sigma_e^2 + \frac{\sigma_e^4 f_s^2}{4}}}$$
$$= \frac{\mathcal{R}_p^2 P_{LO}\bar{P}_{Sp}}{\sqrt{\frac{\pi\gamma_B}{2}\mathcal{R}_p^2 P_{LO}\bar{P}_{Sp}\sigma_e^2 + \frac{\pi^2\gamma_B^2\sigma_e^4}{16}}} \quad (24)$$

When the LO shot noise substantially exceeds the thermal noise, such that $\sigma_e^2 \approx 2q\mathcal{R}_p P_{LO}$, equation (24) simplifies to

$$SNR_p = \frac{\mathcal{R}_p^2 P_{LO}\bar{P}_{Sp}}{\sqrt{q\pi\gamma_B\mathcal{R}_p^3 P_{LO}^2\bar{P}_{Sp} + \frac{q^2\pi^2\mathcal{R}_p^2\gamma_B^2 P_{LO}^2}{4}}}$$
$$= \frac{\mathcal{R}_p\bar{P}_{Sp}}{\sqrt{q\pi\gamma_B\mathcal{R}_p\bar{P}_{Sp} + \frac{q^2\pi^2\gamma_B^2}{4}}} \quad (25)$$

For CW excitation over a total excitation duration $T_{exc}$, the time-domain signal is divided into $N = T_{exc}/T_s$ segments of equal duration $T_s$. Assuming that the spectra obtained from different segments are statistically independent, averaging the $N$ spectra improves the SNR by a factor of $\sqrt{N}$. The peak SNR of the averaged spectrum is therefore

$$SNR'_p = \sqrt{N}SNR_p = \frac{\sqrt{N}\mathcal{R}_p\bar{P}_{Sp}}{\sqrt{q\pi\gamma_B\mathcal{R}_p\bar{P}_{Sp} + \frac{q^2\pi^2\gamma_B^2}{4}}}$$
$$= \frac{\sqrt{\frac{T_{exc}}{T_s}}\mathcal{R}_p\bar{P}_{Sp}}{\sqrt{q\pi\gamma_B\mathcal{R}_p\bar{P}_{Sp} + \frac{q^2\pi^2\gamma_B^2}{4}}} = \frac{A\bar{P}_{Sp}\sqrt{T_{exc}}}{\sqrt{B\bar{P}_{Sp} + C}} \quad (26)$$

where $A = \mathcal{R}_p/\sqrt{T_s}$, $B = q\pi\gamma_B\mathcal{R}_p$ and $C = \frac{q^2\pi^2\gamma_B^2}{4}$ are constants independent of the excitation power. Hereafter, the term $B\bar{P}_{Sp}$, which scales linearly with the Brillouin signal power, is referred to as the Brillouin-signal shot noise term, whereas $C$ which is independent of the excitation power, is referred to as the constant-noise term.

Within this general framework, two distinct operating regimes should be considered separately. First, when the Brillouin-signal shot noise is negligible relative to the constant-noise floor ($B\bar{P}_{Sp} \ll C$), the system operates in the constant-noise-limited regime. Equation (26) then reduces to

$$SNR'_p \propto \bar{P}_{Sp}\sqrt{T_{exc}} \propto P_{exc}\sqrt{T_{exc}} \tag{27}$$

where $P_{exc}$ is the excitation optical power. Thus, the SNR increases linearly with excitation power but only with the square root of excitation duration. Defining an enhancement factor $Q = P_{exc}/P_{ref}$ gives

$$T_{exc} = \frac{T_{ref}}{Q^2} \tag{28}$$

where $P_{ref}$ and $T_{ref}$ denote the excitation power and duration of the reference condition, respectively. For example, at a fixed SNR, increasing the excitation power 53-fold, from $P_{ref} = 530$ m$W$ to $P_{exc} = 28\ W$ ($Q = P_{exc}/P_{ref} = 53$), reduces the required excitation duration from $T_{ref} = 10$ ms to $T_{exc} = 3.6\ \mu s$. The corresponding optical energy is reduced by a factor of $Q$, from 5.3 mJ to 0.1 mJ (Fig. 1b, top).

To mitigate the phototoxicity associated with sustained high-power CW illumination, the shortened effective excitation duration can be distributed over the original acquisition window $T_{ref}$ as a pulse train. The resulting excitation has a peak power $P_{peak} = P_{exc} = QP_{ref}$, a duty cycle $\eta = \frac{T_{exc}}{T_{ref}} = 1/Q^2$ and an average power $P_{ave} = P_{peak}/Q^2$. The SNR under pulsed excitation therefore scales as

$$SNR'_p \propto \sqrt{\frac{P_{peak}^2}{\eta}T_{exc}\eta} = \sqrt{\frac{P_{peak}^2}{Q^2}T_{exc}Q^2}$$
$$= \sqrt{\frac{P_{peak}}{Q^2}P_{peak}T_{ref}} = \sqrt{P_{ave}QP_{ref}T_{ref}} \propto \sqrt{P_{ave}Q} \tag{29}$$

where the final proportionality assumes fixed $P_{ref}$ and $T_{ref}$. Thus, a $Q$-fold increase in peak power permits a $Q$-fold reduction in average power while maintaining the same SNR (Fig. 1b and Supplementary Fig. 9b).

Conversely, when the Brillouin-signal-dependent noise dominates the constant noise floor ($B\bar{P}_{Sp} \gg C$), Eq. (26) reduces to

$$SNR'_p \propto \sqrt{\bar{P}_{Sp}T_{exc}} \propto \sqrt{P_{exc}T_{exc}} \propto \sqrt{P_{ave}T_{ref}} \propto \sqrt{E} \tag{30}$$

where $E = P_{exc}T_{exc} = P_{ave}T_{ref}$ is the total delivered optical energy. In this Brillouin-signal-shot-noise-limited regime, excitation power and duration contribute symmetrically through their product. Increasing the peak power therefore provides no further SNR advantage at a fixed delivered energy. Notably, the term Brillouin-signal shot noise does not describe shot noise generated by directly detecting the fW-scale Brillouin light. It instead refers to the Brillouin-signal-dependent noise term in heterodyne detection. When this term dominates, the SNR depends only on the total delivered energy, recovering the conventional energy-limited scaling of VIPA-based spontaneous Brillouin detection. Although the SNR expression derived here has a mathematical form similar to that in Ref. [2], it is formulated specifically for PHBD. Excess intensity noise from the spontaneous Brillouin signal is neglected because the signal power in biological measurements is typically at the femtowatt level. The previous CW heterodyne Brillouin microscopy study in Ref. [1] reported an approximately inverse dependence of Brillouin-shift precision on excitation power. Within our framework, this scaling is characteristic of the constant-noise-limited regime and indicates dominance of the constant-noise term in the denominator of Eq. (26). The "shot-noise-limited" condition reported in Ref. [1] is therefore more precisely interpreted as LO-shot-noise-dominated detection, rather than Brillouin-signal-shot-noise-limited detection.

Note that coherent time-domain heterodyne detection and pulsed excitation are well established in distributed Brillouin fiber sensing, but their functions and operating regime differ from those of PHBD. In Brillouin optical time-domain reflectometry (BOTDR), the fiber itself is the measurand: optical pulses encode position through time of flight, while the meter-scale scattering length within a kilometer-scale sensing fiber produces a comparatively strong Brillouin return. Such systems can therefore readily approach the Brillouin-signal-shot-noise-limited regime and routinely attain sub-megahertz Brillouin-frequency precision. Because the SNR in this regime is governed primarily by the total number of detected Brillouin photons, redistributing a fixed optical energy into higher peak power provides little intrinsic benefit. PHBD inverts this paradigm in both measurement target and noise physics. In its fiber-probe implementation, the fiber serves as the delivery-and-collection probe rather than the measurand: the signal of interest originates from a micrometer-scale biological volume at the distal facet, while the fiber's own Brillouin response becomes a background. This much shorter interaction length produces substantially weaker biological signals, such that the total detection noise can remain largely independent of excitation power. By identifying and exploiting pulse enhancement in this constant-noise-limited regime, PHBD uses high peak excitation to increase the optical energy efficiency. Thus, rather than employing pulsed heterodyne detection primarily for distributed fiber sensing, PHBD repurposes it as an enabling mechanism for low-dose Brillouin measurements with approximately 10-MHz frequency precision in live-cell microscopy and distal-facet fiber probing.

## Supplementary Note 2. Experimental and simulation-based validation of pulse enhancement.

To validate the principle of pulse enhancement, we compared the experimentally measured Brillouin-shift precision with numerical predictions over a range of peak and average powers at a fixed acquisition time. Shift precision was selected as the primary metric because it directly reflects the measurement performance of PHBD. In the constant-noise-limited regime, the shift precision is inversely proportional to the peak SNR of the measured spectrum and can be approximated as [4]

$$\sigma_{shift} = \frac{1}{SNR'_p} \sqrt{\frac{3}{4} \times \delta \times \gamma_B} \tag{31}$$

where $\delta$ and $\gamma_B$ denote the frequency step and the Brillouin linewidth of the sample,

respectively. The SNR scaling is derived in Supplementary Note 1. The frequency spacing is given by

$$\delta = \frac{f_s}{N_d} = \frac{1}{T_s} \tag{32}$$

where $f_s$ denotes the sampling rate of the digitizer, $T_s$ the acquisition-window duration for each pulse and $N_d = f_s T_s$ denotes the number of digitized samples, and hence FFT points, within each window. Under the assumptions of the model, the effects of $T_s$ on the SNR and frequency spacing compensate, making the predicted shift precision independent of $T_s$.

Considering the laser-system constraints and the required spectral resolution, pulse durations of 14–36 ns were used in PHBD. The number of spectra averaged for each measurement was determined by the pulse repetition rate and total acquisition time. For example, at a repetition rate of 10 kHz and an acquisition time of 10 ms, $N = 100$ pulses were acquired and averaged using the processing workflow described in the Methods. In the simulations, the pulse duration was fixed at 36 ns, corresponding to a frequency spacing of 27.8 MHz. The experimentally measured linewidth of 288 MHz, which includes broadening from both the finite FFT window and the excitation pulse duration, was used in the model.

In the simulations, the LO power was set to 9 mW, matching the experimental condition under which LO shot noise dominated the thermal noise of the BPD (Supplementary Fig. 2). The total detection-noise power spectral density was therefore approximated as

$$\sigma_e{}^2 = 2q\mathcal{R}_p P_{LO} \tag{33}$$

where the responsivity of the BPD was $\mathcal{R}_p = 0.5\ A/W$. The collected Stokes Brillouin power $\bar{P}_{Sp}$ is estimated using a VIPA spectrometer. For an incident excitation power of 1 mW and an effective objective NA of 0.12, $\bar{P}_{Sp}$ was approximately 7.5 fW. This estimation was performed as follows. First, a 1-pW laser beam, generated by attenuating a 100-nW laser with a 50-dB optical attenuator, was directly coupled into the VIPA spectrometer. The camera counts integrated over the laser spot were recorded after subtraction of the mean background. A camera exposure time of 20 ms was used to obtain sufficiently high counts such that pixel-level background fluctuations were negligible. After optimization of the collection efficiency (Methods), a CW laser of controlled power was focused into water, and the collected Brillouin signal was directed to the VIPA spectrometer. Because the Brillouin signal was substantially weaker than the 1-pW calibration beam, a camera exposure time of 200 ms was used. Only the Stokes component was integrated, consistent with the Stokes-only detection used in PHBD. The collected Brillouin power was then estimated by comparing the background-subtracted integrated camera counts of the Brillouin and calibration measurements, with corrections for their different exposure times and incident excitation powers.

Parameters corresponding to the first experimental point were used as a baseline, with a peak power $P_{b,peak} = 28\ W$, pulse repetition rate $R_b = 10\ kHz$, average power $P_{b,ave} = 10\ mW$, exposure time $T_b = 10\ ms$ and averaging number $N_{b,ave} = R_b \times$ $T_b = 100$. For a simulated peak power $P_{s,peak}$, we define the enhancement factor as $Q = P_{s,peak} / P_{b,peak}$. The repetition rate was then set to $R_s = R_b / Q^2$, giving an average power $P_{s,ave} = P_{b,ave}/Q$. The total acquisition time was held fixed at $T_b$, such that the number of averaged spectra became $N_{s,ave} = N_{b,ave}/Q^2$.

The simulated and experimental results are shown in Supplementary Fig. 9. The model predicts that the Brillouin-signal shot noise variance equals the constant-noise variance at a peak excitation power of approximately 9.7 W. Above this crossover, the benefit of increasing the peak power progressively saturates, consistent with the experimental measurements. At a peak power of approximately 19.3 W, the Brillouin-signal shot noise variance is twice the constant-noise variance, which we use as the criterion for entering the Brillouin-signal-shot-noise-limited regime. In this regime, the SNR scales with the square root of the Brillouin signal power, such that increasing the peak power and proportionally extending the acquisition time provide equivalent SNR improvements.

Because the Fabry–Pérot filter limited the experimentally accessible CW excitation power to below 500 mW, the 100%-duty-cycle limit of our system was estimated from the simulation. The modest discrepancy between simulation and experiment may arise from the reduced validity of the assumed inverse relationship between shift precision and SNR (Eq. (31)) as the system departs from the constant-noise-limited regime at higher peak powers (Supplementary Fig. 9), together with the frequency-dependent BPD noise floor, the non-flat electronic frequency response (Supplementary Figs. 2, 3 and 8), and uncertainty in the measured Brillouin scattering efficiency of water.

## Supplementary Note 3. Performance comparison between pulsed and CW excitation in heterodyne detection

We compared pulsed and CW excitation at two levels: within our heterodyne detection system and against a previously reported CW heterodyne implementation [1]. Within our system, pulse enhancement reduced the average power required to achieve a Brillouin-shift precision below 10 MHz at the same 10-ms acquisition time by 53-fold, from a modeled CW requirement of 527 mW (rounded to 530 mW in Fig. 1b) to 10 mW under pulsed excitation. As the peak power increased, the pulse-enhancement benefit approached its maximum owing to the increasing contribution of Brillouin-signal shot noise. Accordingly, the Brillouin-shift precision decreased from 5.5 MHz to 9.9 MHz at the lowest-average-power operating point, while remaining sufficient for biological imaging.

The previously reported CW heterodyne implementation achieved a Brillouin-shift precision of approximately 9 MHz in water using 345 mW of excitation power and a 10-ms acquisition time [1]. At the same acquisition time and comparable precision, PHBD required only 10 mW of average power, corresponding to an approximately 35-fold reduction. The difference between this cross-system enhancement factor and the 53-fold within-system value likely reflects differences in system implementation and excitation wavelength. Notably, for biological imaging, the previous implementation used 276 mW with a 100-ms pixel dwell time to image HeLa cells, whereas free-space PHBD used 30 mW with a 10-ms pixel dwell time to image strongly scattering plant and animal tissues. These operating conditions correspond to a 92-fold reduction in delivered optical energy per pixel, substantially lowering optical exposure and the associated risk of phototoxicity.

## Supplementary Note 4. Verification of spectral accuracy and fidelity

We verified the absolute frequency accuracy of the spectrometer by comparing the beat frequency between two single-frequency CW lasers, measured with a frequency counter, against the value retrieved by PHBD. The two measurements agreed to within 0.1 MHz, demonstrating high frequency accuracy.

Next, to further validate spectral fidelity, we performed side-by-side measurements at varying numerical apertures (NAs) using both free-space PHBD and an SBS microscope configured according to Ref. 12. The Brillouin linewidths measured with both low- and high-NA objectives were consistent between the two systems within experimental uncertainty (Supplementary Fig. 11). Notably, PHBD intrinsically avoids the spectral broadening induced by lock-in amplifiers, a common broadening factor in SBS systems.

## Supplementary Note 5. Quantitative comparison of PHBD with other Brillouin techniques

We quantitatively benchmarked PHBD against representative Brillouin implementations. To facilitate direct comparisons, we operated PHBD at selected combinations of average excitation power and acquisition time and measured the resulting Brillouin-shift precision in water (Supplementary Fig. 16). At an average excitation power of 30 mW and an acquisition time of 3 ms, PHBD achieved a shift precision of 9.0 MHz (Supplementary Fig. 16a). Compared with previously reported confocal Brillouin microscopy approaches, PHBD reached comparable shift precision with approximately 3.3-fold lower optical energy exposure under the evaluated conditions [5–7]. When matched to orthogonal line-scanning Brillouin microscopy in both acquisition time (1 ms) and shift precision (18.6 MHz), PHBD required a comparable excitation power of 20 mW (Supplementary Fig. 16c) [8,9]. When matched to full-field Brillouin microscopy in shift precision, PHBD achieved an approximately twofold shorter acquisition time of 0.01 ms while using approximately twice the excitation power (160 mW; Supplementary Fig. 16d) [10].

Beyond these spontaneous Brillouin implementations, PHBD achieved a tenfold shorter acquisition time than pulse-chopped stimulated Brillouin scattering microscopy at comparable excitation power and shift precision (Supplementary Fig. 16e) [11]. Recently developed pulsed-laser SBS approaches achieve acquisition times approximately one order of magnitude shorter than PHBD [12]. However, they retain a dual-objective transmission geometry that restricts measurements to optically semi-transparent samples and offer lower spectral resolution than PHBD.

Overall, these comparisons highlight that PHBD provides a combination of high shift sensitivity, spectral resolution and flexible optical access. In particular, its single-ended and fiber-compatible architecture expands Brillouin measurements toward geometries and samples that are challenging for conventional transmission-based implementations.

## Supplementary Note 6. Compatibility of Brillouin spectroscopic techniques with endoscopic measurement

The opposing dual-objective geometry required for stimulated Brillouin microscopy and the orthogonal geometries used in line-scanning and full-field spontaneous Brillouin microscopy preclude endoscopic implementation because they require optical access from multiple directions. Single-ended spontaneous Brillouin approaches are more directly compatible with fiber delivery and collection. Here, we evaluate the integration of three distinct spectrometer architectures into endoscopic platforms: CW-heterodyne, VIPA, and our proposed pulsed-heterodyne design.

In conventional endoscopic systems, the excitation laser and signal light are co-propagated through a single optical fiber, causing intrinsic fiber-generated signals (including spontaneous Brillouin and Raman scattering) to superimpose on the sample's Brillouin signal at the detector. In our fiber-coupled VIPA test, Brillouin scattering generated in a 0.5-m length of PM780-HP fiber saturated the sCMOS camera within the spectral region of interest at 45 mW incident power and a 50-ms exposure. Such saturation risks irreversible damage to high-sensitivity sCMOS or EMCCD cameras. In addition, the finite free spectral range (FSR; typically, ~30 GHz) of the VIPA causes the high-frequency fiber background to alias directly into the biological signal band. When the incident power is reduced to prevent camera saturation and the resulting fiber background is subtracted, the residual noise can overwhelm the weak Brillouin signal from the target tissue, ultimately degrading image quality.

In heterodyne detection, high-frequency noise is rejected by the balanced photodetector, whose detection bandwidth acts as a natural low-pass filter. For fiber-coupled measurements operating in the Brillouin-signal-shot-noise-limited regime, the lower collection efficiency relative to the free-space configuration required approximately twice the delivered optical energy to achieve comparable Brillouin-shift precision in water (Supplementary Fig. 5c–e). Applying this factor to the operating point reported for the previous CW heterodyne implementation gives estimated conditions of approximately 390 mW at 100 ms or 276 mW at 200 ms. In either case, the resulting thermal load increases the risk of phototoxicity, limiting the suitability of this approach for *in vivo* imaging of biological tissues.

Pulsed-heterodyne detection addresses both challenges simultaneously: heterodyne detection intrinsically suppresses fiber-generated background signals, while pulsed excitation improves energy efficiency, maintaining imaging performance at reasonable incident powers and dwell times (Supplementary Fig. 5c–e). Together, these properties make the pulsed-heterodyne architecture an appropriate platform for endoscopic Brillouin spectroscopy and imaging.

## Supplementary Note 7. Objective selection for different applications

To optimize performance across different experimental conditions, we mainly used three objective configurations. A low-numerical-aperture (NA) objective (ACHN10XP, Olympus) was used primarily to characterize the intrinsic linewidth of water and assess the performance of the system. The incident beam diameter was 4.4 mm, smaller than the 9-mm back aperture of the objective, yielding an effective NA of 0.12. A second objective with a moderately higher effective NA (ACHN20XP, Olympus) was used to compare water linewidths measured with the SBS microscope and the free-space PHBD configuration. Because its back aperture was also underfilled by the incident beam, the effective NA was 0.24. For imaging cultured cells (Supplementary Fig. 18), plant tissues and porcine kidney tissue, higher spatial resolution was required to resolve subcellular structures. We therefore used a high-NA objective (LUCPLFLN60X, Olympus). In this configuration, the incident beam filled the back aperture, enabling use of the full NA of 0.7 and providing a measured spatial resolution of 0.6 × 0.6 × 2.4 $\mu m^3$.

## Supplementary Note 8. Limitations and further developments

The current fiber-coupled PHBD has two main limitations. First, the lensed probe used in the present implementation remains relatively large, with its size mainly determined by the two-lens assembly (Supplementary Fig. 5b). Future implementations could use shorter-focal-length, higher-NA aspherical lenses to reduce the probe diameter to approximately 3 mm without compromising spatial resolution. The shorter focal length would also allow the fiber end to be positioned closer to the optics, reducing the overall probe length to below 15 mm.

Second, the current lensed probe does not yet incorporate intrinsic beam-scanning capability. Therefore, the porcine gastric tissue experiment was performed *ex vivo*, with lateral scanning provided by an external piezostage rather than by the probe itself. Piezoelectric benders are promising candidates because they support distal scanning, which is well suited to the pixel acquisition required for Brillouin imaging. Typical piezoelectric benders can provide a distal scanning range of several hundred micrometers. By mounting two piezobenders orthogonally, two-axis beam scanning could be implemented within the probe. Further miniaturization of the optical probe and integration of two-axis distal scanning would reduce probe invasiveness, improve compatibility with clinically relevant endoscopic channels and enable localized mechanical mapping in tissues that cannot be externally scanned.

Another possible improvement concerns the heterodyne spectrometer. In the present implementation, only the Stokes Brillouin signal is detected. The usable Brillouin signal could therefore be increased by additionally detecting the anti-Stokes component. For example, a carrier-suppressed electro-optic modulator could be used to generate a double-sideband local oscillator, enabling simultaneous heterodyne detection of the Stokes and anti-Stokes Brillouin sidebands. Such a configuration could, in principle, nearly double the detected Brillouin signal, thereby doubling the acquisition speed.

### Supplementary References